\documentclass[twocolumn,pre,amssymb,amsmath,natbib,showpacs,floatfix]{revtex4}%
\usepackage[dvips]{graphicx}
\usepackage{color}
\usepackage[final]{showkeys}
\newcommand{\Eref}[1]{{Eq.~(\ref{#1})}}
\newcommand{\Fref}[1]{{Fig.~\ref{#1}}}
\newcommand{\etal}{{\it et al.}}

\begin{document}

\title{Nearest-Neighbor Distributions and Tunneling Splittings in
  Interacting Many-Body Two-Level Boson Systems}%

\author{Sa\'ul Hern\'andez-Quiroz}%
\email{saul@cicc.unam.mx}%
\affiliation{%
  Instituto de Ciencias F\'{\i}sicas, Universidad Nacional Aut\'onoma
  de M\'exico (UNAM), 62210 Cuernavaca, Morelos, Mexico}%
\affiliation{%
  Facultad de Ciencias, Universidad Aut\'onoma del Estado de Morelos
  (UAEM), 62209-Cuernavaca, Morelos, Mexico }%
\author{Luis Benet}%
\email{benet@fis.unam.mx}%
\affiliation{%
  Instituto de Ciencias F\'{\i}sicas, Universidad Nacional Aut\'onoma
  de M\'exico (UNAM), 62210 Cuernavaca, Morelos, Mexico}%

\date{\today}

\begin{abstract}
  We study the nearest-neighbor distributions of the $k$-body embedded
  ensembles of random matrices for $n$ bosons distributed over
  two-degenerate single-particle states. This ensemble, as a function
  of $k$, displays a transition from harmonic oscillator behavior
  ($k=1$) to random matrix type behavior ($k=n$). We show that a large
  and robust quasi-degeneracy is present for a wide interval of
  values of $k$ when the ensemble is time-reversal invariant. These
  quasi-degenerate levels are Shnirelman doublets which appear due to
  the integrability and time-reversal invariance of the underlying
  classical systems. We present results related to the frequency in
  the spectrum of these degenerate levels in terms of $k$, and discuss
  the statistical properties of the splittings of these doublets.
\end{abstract}

\maketitle


\section{Introduction}

The theoretical and experimental understanding of interacting
many-body quantum systems has undergone considerable development in
recent years. First, random matrix theory (RMT) has been quite
successful in describing the statistical properties of the
fluctuations of the spectra of complex quantum systems, which include
many-body interacting systems. Examples range from nuclear physics to
disordered systems, including elasto-mechanical vibrations and quantum
analog systems to classical chaotic billiards (see~\cite{GMGW1998} for
a detailed review). While this modeling has been quite successful, RMT
is not a realistic theory since it assumes many-body forces between
the constituents. More realistic stochastic model considering $k$ body
interactions are the embedded ensembles, initially introduced by Mon
and French~\cite{MonFrench}. This model can be defined for fermions
and bosons~\cite{BW2003}, and may be viewed as the generic models for
stochasticity in many-body systems.

Second, ultra-cold bosonic gases confined in optical lattices have
become quite important due to the relatively simplicity to handle
these systems experimentally~\cite{MorschOberthaler2006}. In
particular, Bose-Einstein condensates (BECs) in a double-well
potential is a common object of study~\cite{GatiOberthaler2007}. This
system exhibits a great variety of interesting quantum phenomena, such
as interference~\cite{Andrews1997}, tunneling and
self-trapping~\cite{Milburn1997,Albiez2005}, Josephson
oscillations~\cite{Cataliotti2001}, and
entanglement~\cite{Micheli2003}.

From the theoretical point of view, the two-level bosonic systems have
been addressed using the mean field treatment of the Gross-Pitaevski
equation~\cite{Milburn1997,Franzosi2001}, and the two-site
Bose-Hubbard Hamiltonian. The latter can be written
as~\cite{Zwerger2003,Bloch2005}
\begin{eqnarray}
  \label{BoseHubbard}
  H_{\rm BH} & = & \delta(n_1-n_2) 
  - J(\hat b_1^\dagger \hat b_2+\hat b_2^\dagger \hat b_1) \nonumber\\
  & & + \frac{U}{2}[ n_1(n_1-1)+n_2(n_2-1)]\, .
\end{eqnarray}
Here, $\hat b_i^\dagger$ and $\hat b_i$ are creation and annihilation
operators for a boson on the $i$th site ($i=1,2$) and $n_i = \hat
b_i^\dagger \hat b_i$ is the total number of bosons on that level,
$\delta$ is the energy difference of one-boson energies among the two
sites, $U$ is the on-site two-body interaction strength, and $J$ is the
hoping or tunneling parameter. The two-mode approximation in
\Eref{BoseHubbard} is valid as long as the interaction energy $U$ is
much smaller than the level spacing of the external
trap~\cite{Milburn1997}.

The experimental observation of macroscopic tunneling of bosons in a
double well when the initial difference of population is below a
critical value~\cite{Albiez2005}, predicted in
Ref.~\cite{Milburn1997}, can be understood from the spectral
properties of \Eref{BoseHubbard}. For the simpler case $\delta=0$, the
spectrum consists of a lower region of nearly equidistant levels and
an upper one displaying nearly degenerate doublets. The latter are
actually responsible for the suppression of tunneling; it has also
been shown that coherences among nearby doublets yield oscillations
with very small amplitude~\cite{Salgueiro2007}. Taking the
semiclassical limit, the system has a phase space representation
similar to a pendulum, with the almost equidistant levels being
associated with the libration zone and the nearly degenerate
levels with the rotation zones.

In this paper, we study the statistical properties of the spectrum of
$n$ bosons distributed on two levels coupled through random $k$-body
interactions. Thus, we merge the successful stochastic modeling of RMT
with systems of the form of \Eref{BoseHubbard}. This ensemble is
actually a generalization of the Bose-Hubbard type of Hamiltonians, in
particular with respect to the range of the interaction $k$. Each
member of the ensemble is Liouville integrable (independently of $k$)
in the classical limit~\cite{BJL2003}. Yet, the spectral statistics of
the ensemble correspond to a picket fence for $k=1$, and follow RMT
predictions for $k=n$~\cite{BLS2003}. These facts make the ensemble
somewhat special: completely integrable systems are associated with
Poisson statistics, which is known as the Berry-Tabor
conjecture~\cite{BerryTabor1977}. In addition, the spectral
fluctuations of classically fully chaotic systems typically follow RMT
predictions, which is known as the Bohigas-Giannoni-Schmit (or
quantum-chaos) conjecture~\cite{quantumchaos}. We shall thus study the
transition in the spectral statistics in terms of $k$, considering the
nearest-neighbor distribution as well as the occurrence and statistics
of tunneling splittings. We shall address this for the case when the
ensemble is time-reversal invariant ($\beta=1$) or when this symmetry
is broken ($\beta=2$). We find a systematic appearance of
quasi-degeneracies on a large interval of $k$ for the time-reversal
case, which points out the underlying integrability properties of the
members of the ensemble due to a theorem by
Shnirelman~\cite{Shnirelman1975,Chi1995}. Moreover, the number of such
doublets as well as the statistics of the associated splittings
display a dependence upon $k$. These results may be interesting for
the understanding and modeling of three-body interactions in cold
gases~\cite{Buechler2007}.

The paper is organized as follows. In Sec.~\ref{Sec:k-body}, we
present the $k$-body embedded ensembles of random matrices for
two-level boson systems, and review some important properties of this
ensemble. In Sec.~\ref{Sec:SpecStat}, we discuss the nearest-neighbor
distribution of the ensembles in terms of the interaction parameter
$k$ for both cases of Dyson's parameter $\beta$. We obtain the
systematic appearance of quasi-degenerate states in the spectrum
linked to the $\beta=1$ case, and address the dependence of their
number with respect to $k$. In Sec.~\ref{Sec:PhaseSpace}, we present
the semiclassical limit of this ensemble and describe the structure of
the corresponding phase space. Section~\ref{Sec:Splittings} is devoted
to the identification of the $\beta=1$ quasi-degenerate states and
present results on the statistical properties of their spacings. In
Sec.~\ref{Sec:Concl}, we present a summary of our results and the
conclusions.

\section{$k$-body interacting two-level boson ensemble}
\label{Sec:k-body}

We begin defining the most general $k$-body interaction of $n$
spin-less bosons distributed in two single-particle levels which, for
simplicity, are assumed to be degenerate [case $\delta=0$ in
\Eref{BoseHubbard}]. The single-particle states are associated with
the operators $\hat b_j^\dagger$ and $\hat b_j$, with $j=1,2$, which,
respectively, create or annihilate one boson on the single-particle
level $j$. These operators satisfy the usual commutation relations for
bosons. The normalized $n$-boson states are specified by
$|\mu_r^{(n)}\rangle = ({\cal N}_r^{(n)})^{-1} (\hat b_1^\dagger)^r (\hat
b_2^\dagger)^{n-r}|0\rangle$, where ${\cal N}_r^{(n)}=[r!
(n-r)!]^{1/2}$ is a normalization constant and $|0\rangle$ is the
vacuum state. The Hilbert--space dimension is $N=n+1$.  In
second-quantized form, the most general Hamiltonian $\hat
H_k^{(\beta)}$ with $k$-body interactions can be written
as~\cite{Asaga2001}
\begin{equation}
  \label{eq1}
  {\hat H_k^{(\beta)}} = \sum_{r,s=0}^k \, v_{r,s}^{(\beta)} \,
  \frac{ (\hat b_1^\dagger )^{r} ( \hat b_2^\dagger )^{k-r} 
    (\hat b_1 )^{s} ( \hat b_2 )^{k-s} }{{\cal N}_r^{(k)} {\cal N}_s^{(k)}}  \ .
\end{equation}
Physically, ${\hat H_k^{(\beta)}}$ in \Eref{eq1} corresponds to $n$
bosons confined, e.g., in a double---well potential, coupled only
through $k$-body interactions. Clearly, the degenerate Bose-Hubbard
model \Eref{BoseHubbard} is a particular choice of the parameters for
the combination $ {\hat H_{k=1}^{(1)}}+ {\hat H_{k=2}^{(1)}}$.

Stochasticity is built into the Hamiltonian $\hat H_k^{(\beta)}$ at
the level of the $k$-body matrix elements $v_{r,s}^{(\beta)}$. These
matrix elements are assumed to be Gaussian distributed independent
random variables with zero mean and constant variance $v_0^2=1$. Then,
$\overline{ v_{r,s}^{(\beta)} v_{r',s'}^{(\beta)}} = v_0^2 (
\delta_{r',s} \delta_{r,s'} + \delta_{\beta,1} \delta_{r,r'}
\delta_{s,s'} )$, where the over line denotes ensemble average. As in
the case of the canonical random matrix ensembles~\cite{GMGW1998},
Dyson's parameter $\beta$ distinguishes the symmetry properties with
respect to time-reversal invariance: $\beta=1$ corresponds to the case
where time-reversal symmetry holds while the broken time-reversal case
is denoted by $\beta=2$. The $k$-body interaction matrix $v^{(\beta)}$
is thus a member of the Gaussian orthogonal ensemble (GOE) for
$\beta=1$ or Gaussian unitary ensemble (GUE) for $\beta=2$. This
defines completely the $k$-body embedded ensemble of random matrices
for bosons distributed in $l=2$ levels. Without loss of generality, in
the following, we set $v_0=1$. The combinatorial factors ${\cal
  N}_r^{(k)}$ in \Eref{eq1} are actually introduced in order to have
{\it an exact identity} of the embedded ensembles with the canonical
ensembles of RMT when $k=n$~\cite{BW2003,Asaga2001}. Indeed, the
factors ${\cal N}_r^{(k)}$ cancel the square-root factors that appear
by operating the $k=n$ creation and annihilation operators onto the
many-body states $|\mu_r^{(n)}\rangle$. Then, the central-limit
theorem implies that the matrix elements become independent Gaussian
distributed random variables; this is precisely the definition of the
canonical ensembles of RMT. Consequently, for $k=n$, all spectral
fluctuations correspond exactly to the predictions of RMT.

By construction, the number operator $\hat n=\hat b_1^\dagger\hat b_1 +
\hat b_2^\dagger\hat b_2$ commutes with the Hamiltonian $\hat
H_{k}^{(\beta)}$ for all values of the rank of the interaction
$k$. The Hamiltonian is thus block diagonal in the occupation-number
basis $|\mu^{(n)}\rangle$ defined above. For a given value $k$, the
number of independent random variables of the ensemble is
$K_\beta(k)=\beta(k+1)(k+1+\delta_{\beta,1})/2$, which in general is
smaller than the Hilbert--space dimension $N=n+1$. Therefore, for $k\ll
n$, the matrix elements of the Hamiltonian $\hat H_{k}^{(\beta)}$ are
correlated, i.e., the number of independent matrix elements of the
Hamiltonian is larger than the number of independent random
variables. Moreover, some matrix elements are identically zero.

\section{Spectral statistics in terms of $k$}
\label{Sec:SpecStat}

The evaluation of statistical measures of the spectrum requires
unfolding the spectra, which removes the non-universal
system-dependent contributions. This can be done by performing the
unfolding individually for each spectrum (spectral unfolding) or by a
single transformation used for all members of the ensemble (ensemble
unfolding). In the context of spectral fluctuations, ergodicity
implies that the results are independent of the unfolding method. 

\begin{figure*}
  \includegraphics[width=8cm]{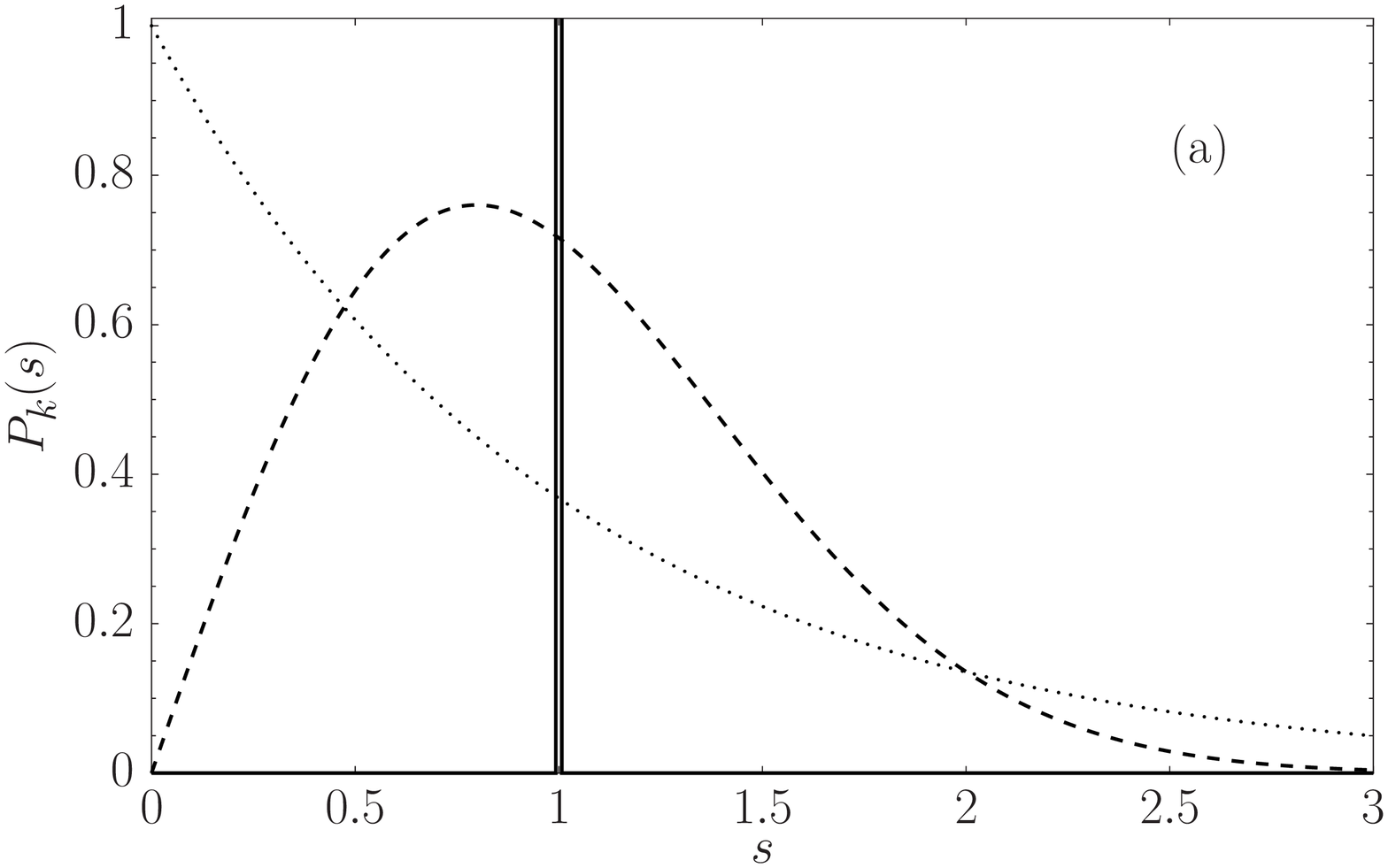}
  \includegraphics[width=8cm]{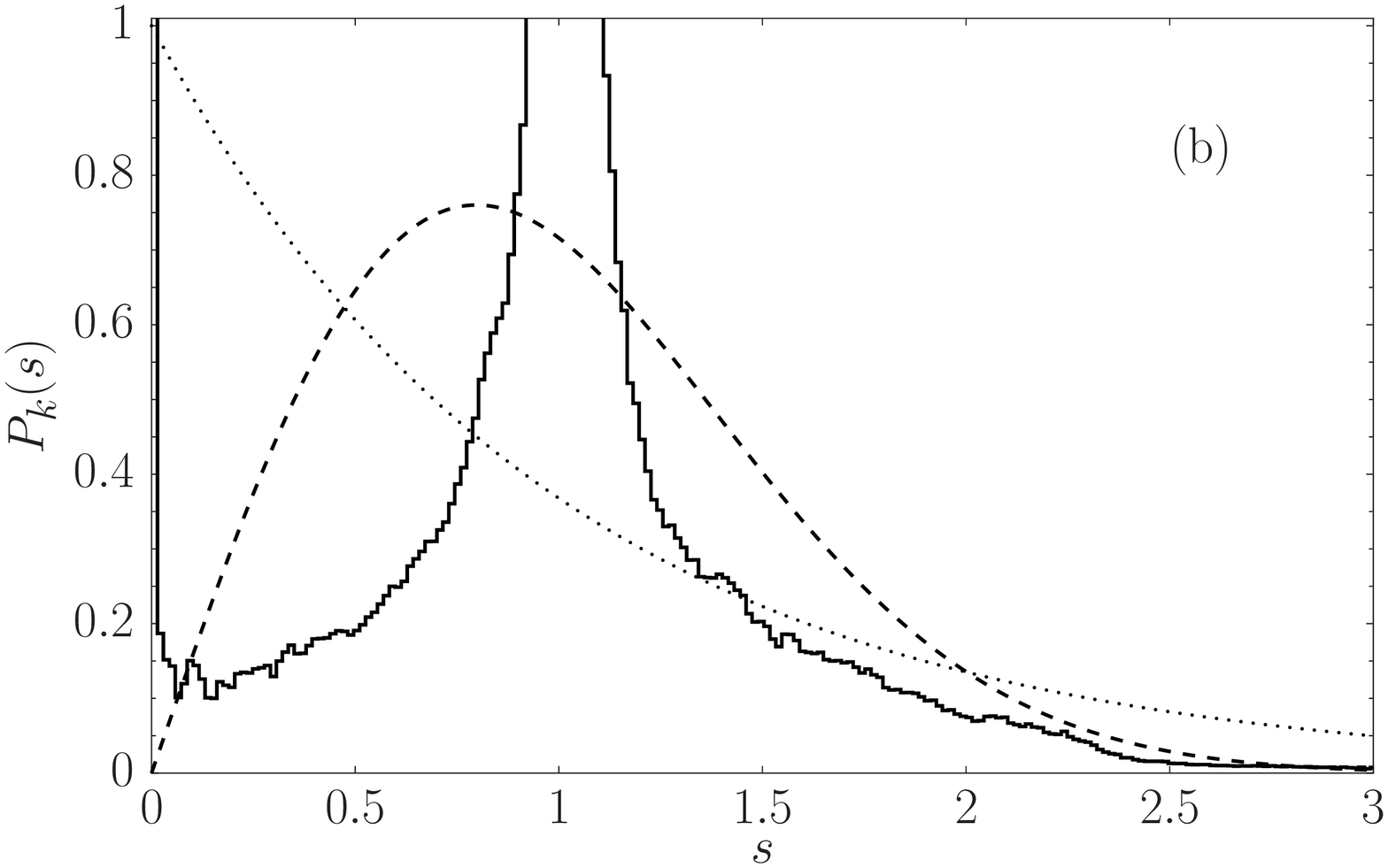}
  \includegraphics[width=8cm]{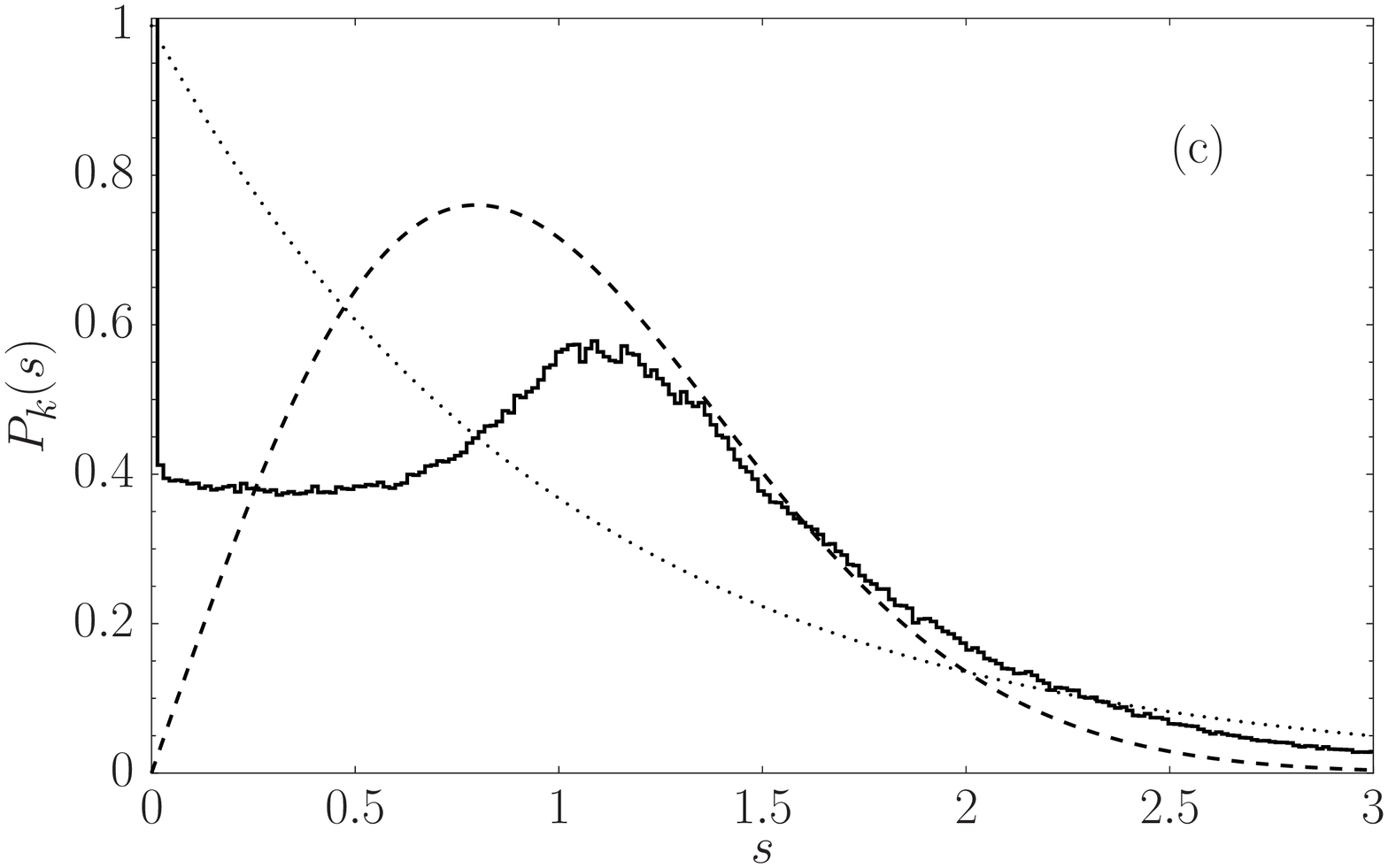}
  \includegraphics[width=8cm]{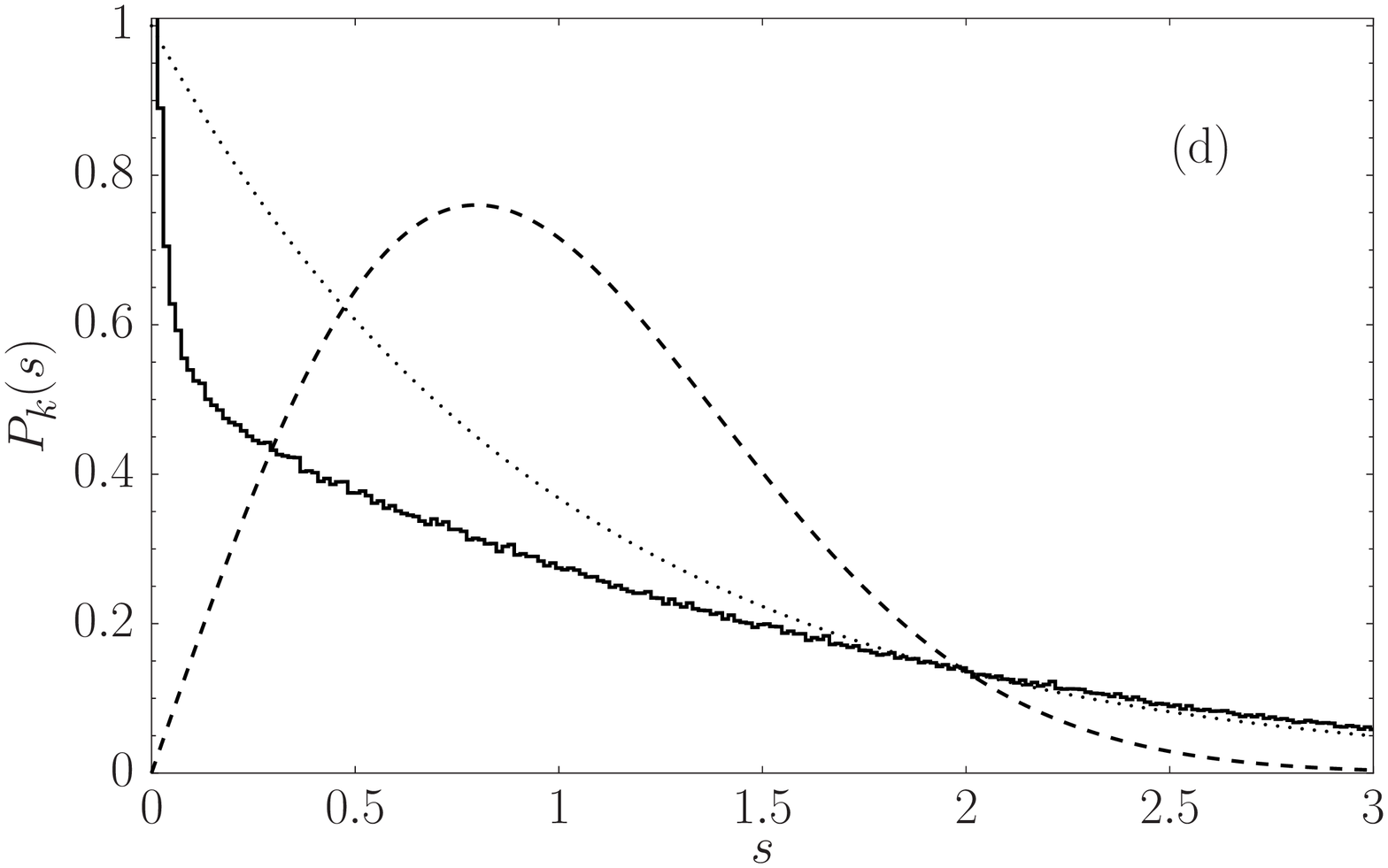}
  \includegraphics[width=8cm]{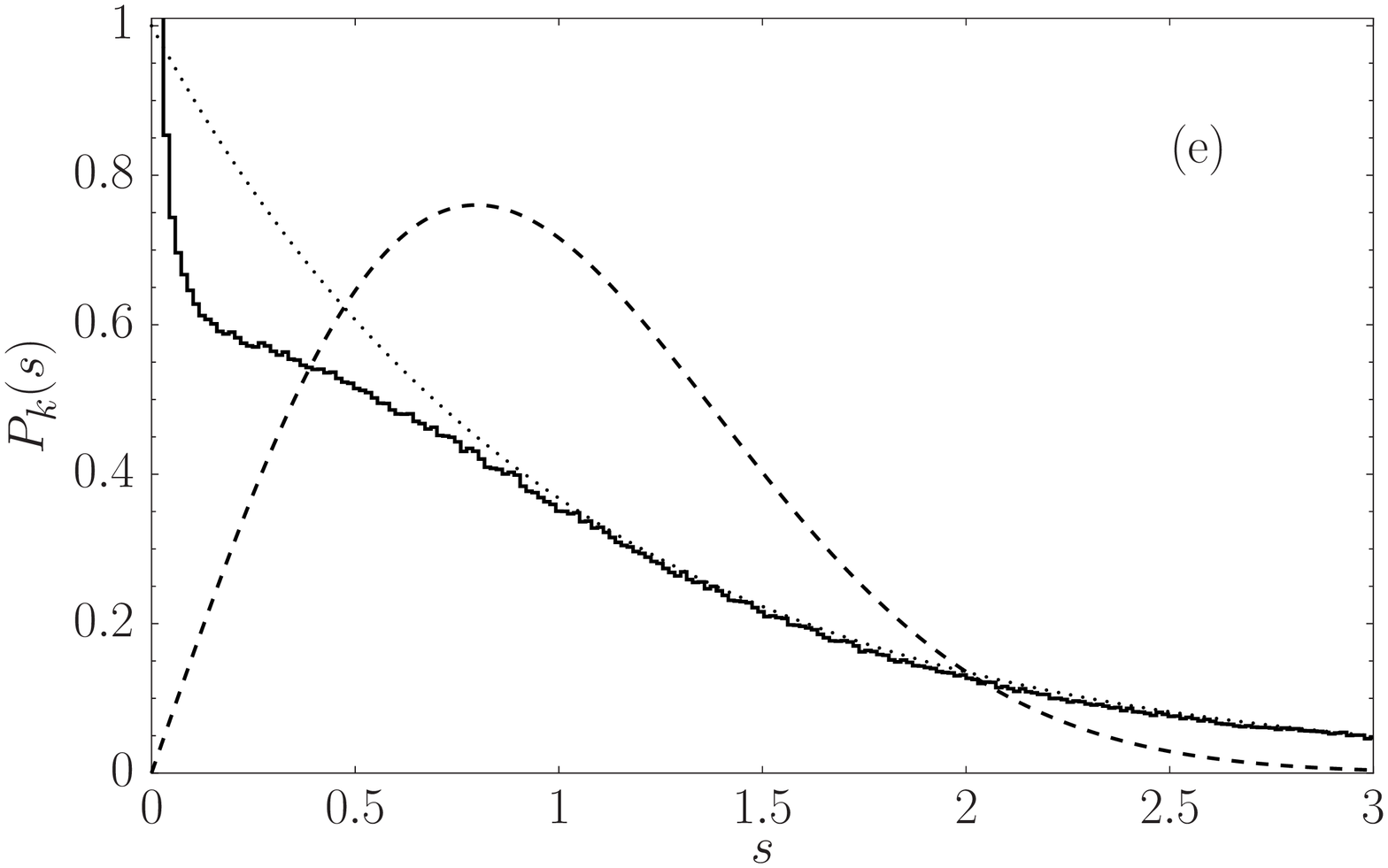}
  \includegraphics[width=8cm]{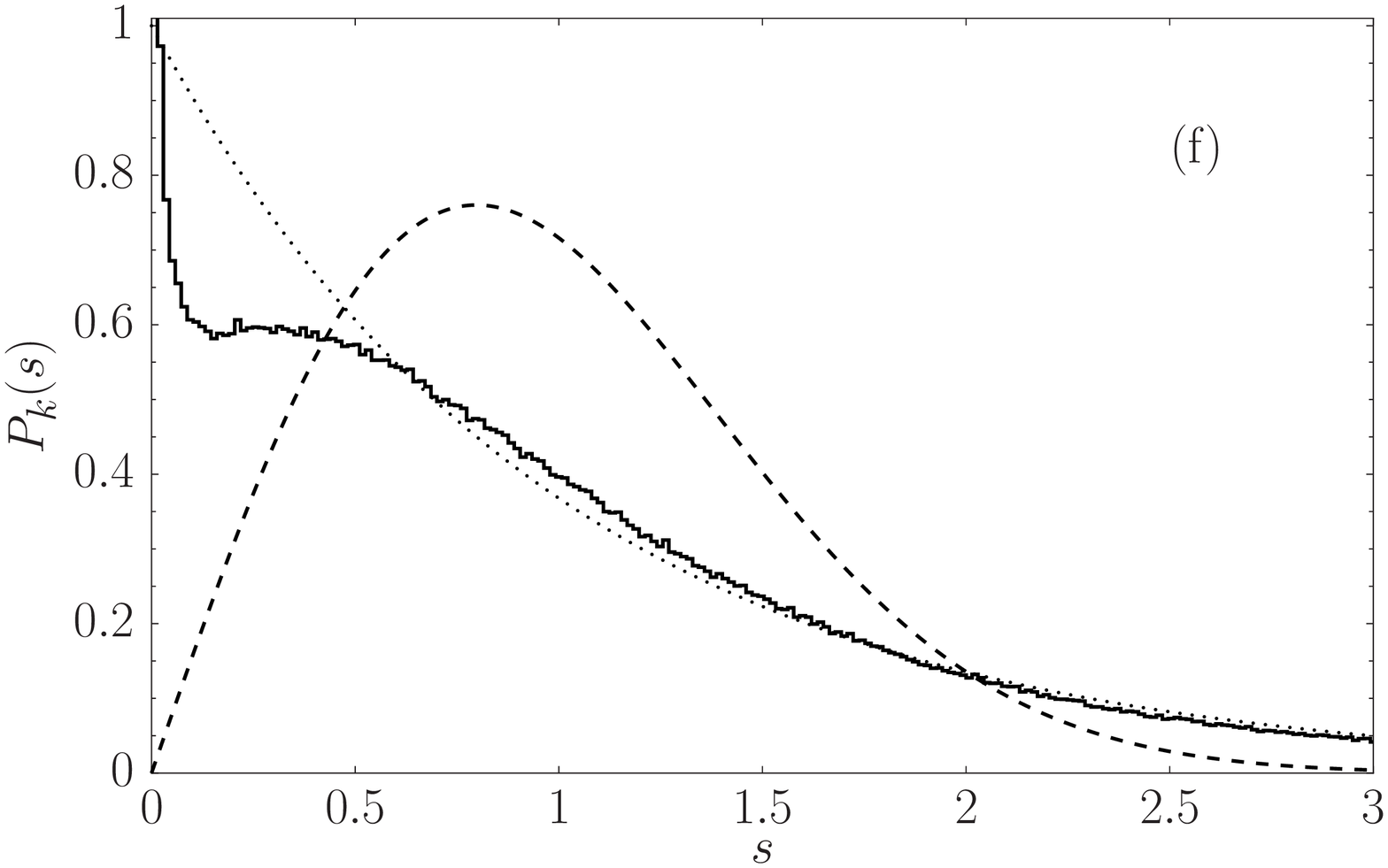}
  \includegraphics[width=8cm]{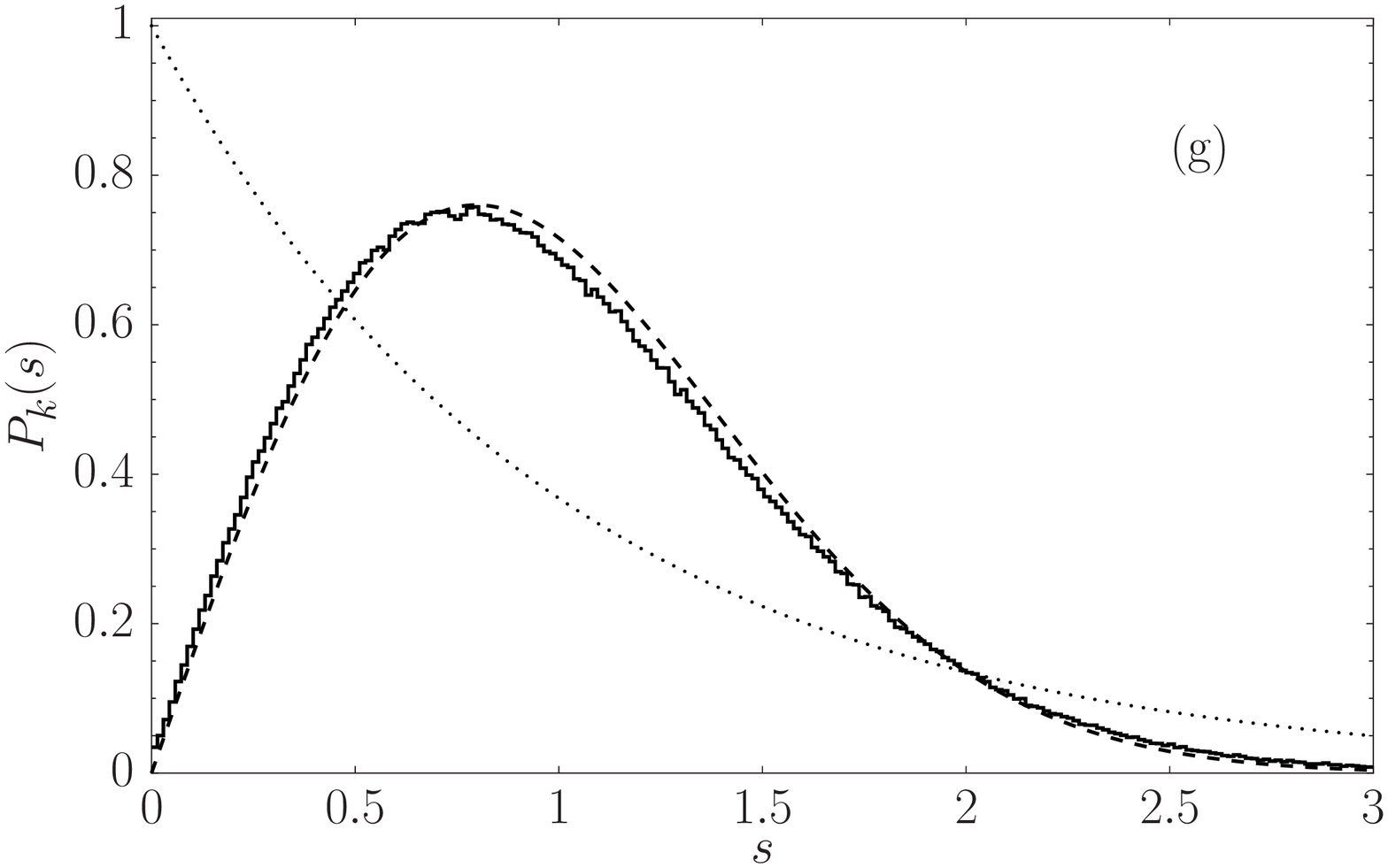}
  \includegraphics[width=8cm]{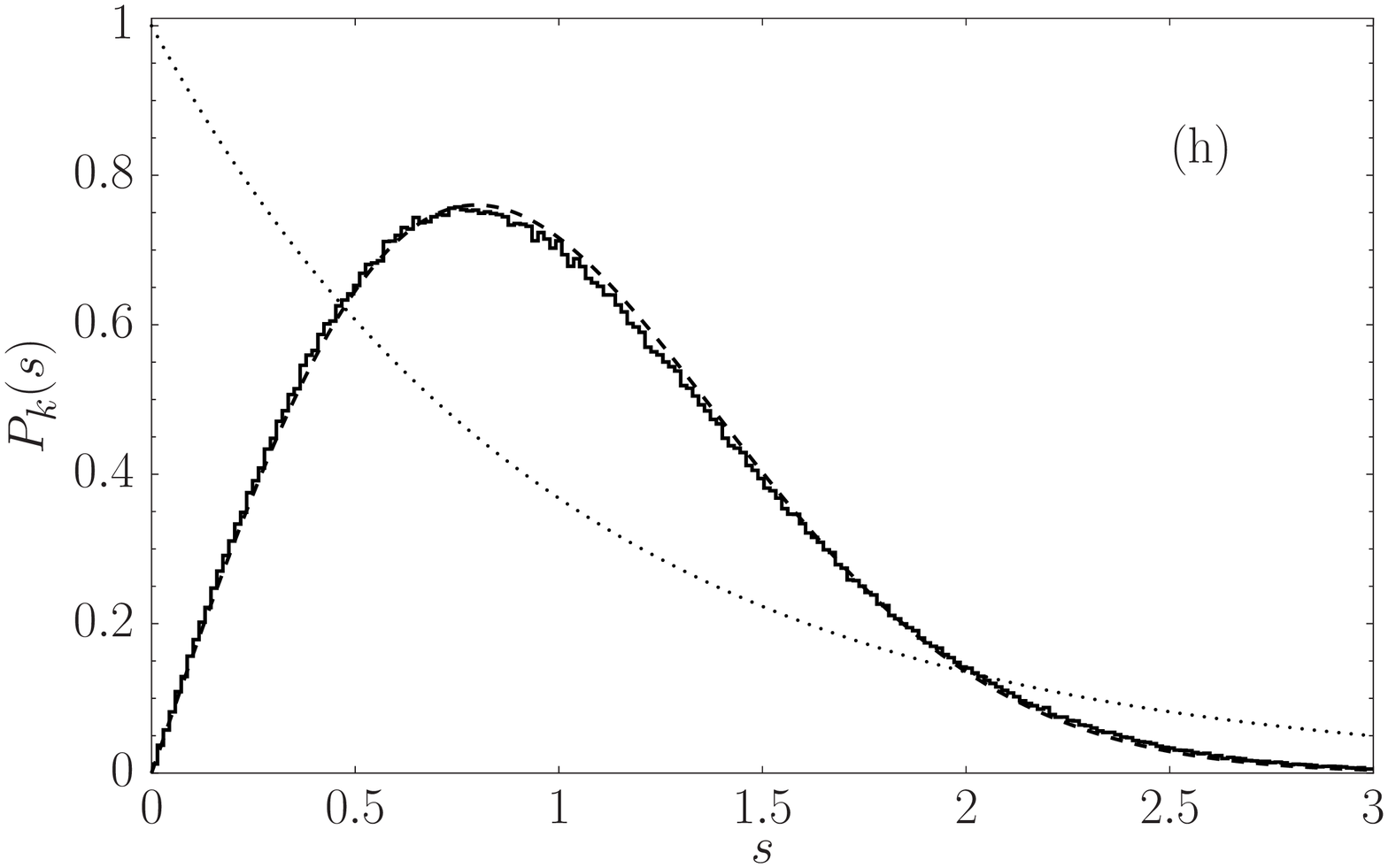}
  \caption{ Nearest-neighbor distribution $P_k(s)$ for the $k$-body
    interacting two-level boson ensemble for $\beta=1$, $n=2000$ and
    (a)~$k=1$, (b)~$k=2$, (c)~$k=10$, (d)~$k=200$, (e)~$k=1000$,
    (f)~$k=1150$, (g)~$k=1850$ and (h)~$k=2000$. Notice the large peak
    observed at $s=0$, which is linked with the occurrence of
    quasi-degenerate levels. The dashed curve corresponds to the
    Wigner surmise for $\beta=1$, while the dotted curve is the
    Poisson distribution.}%
  \label{fig1}%
\end{figure*}

\begin{figure*}
  \includegraphics[width=8cm]{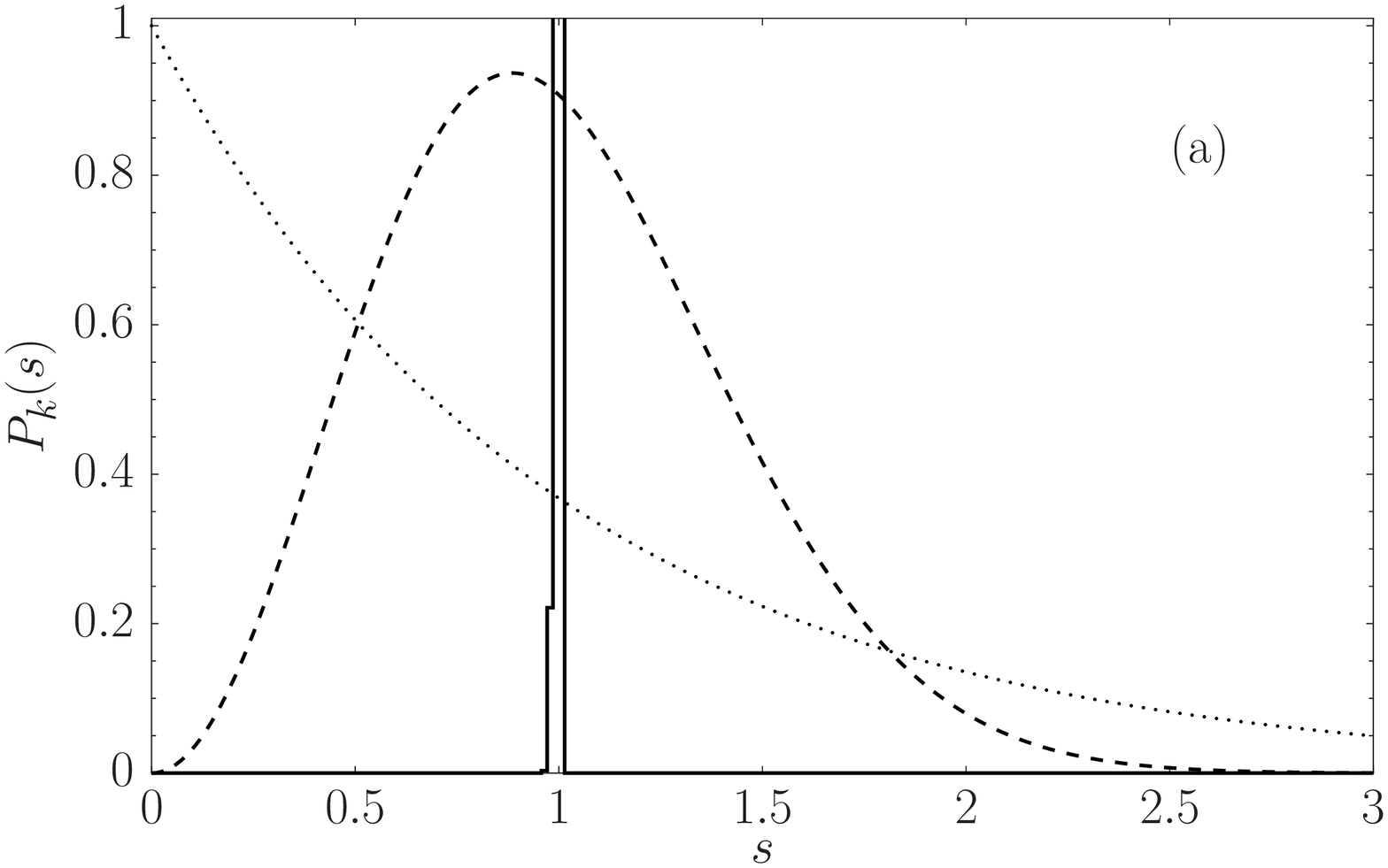}
  \includegraphics[width=8cm]{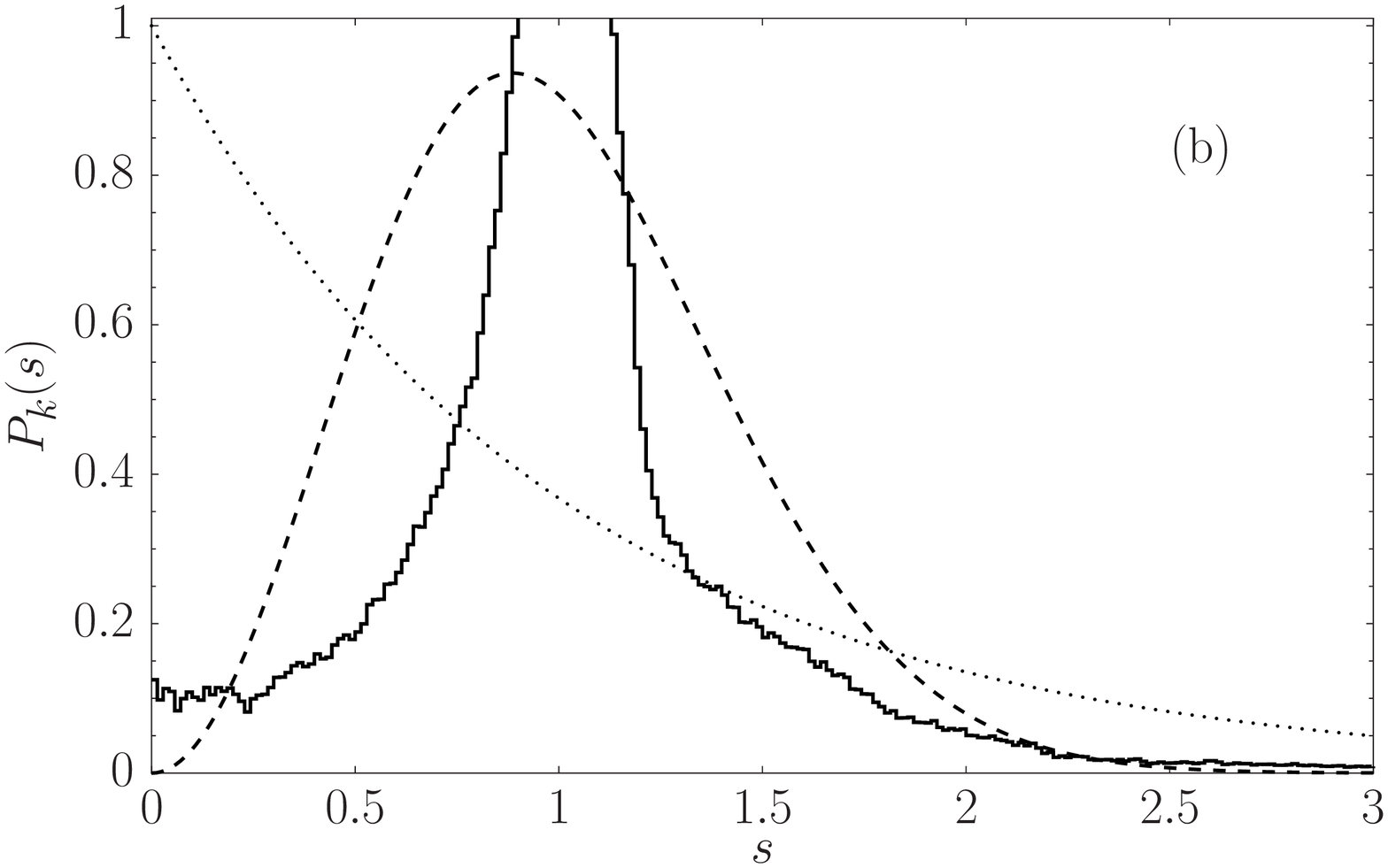}
  \includegraphics[width=8cm]{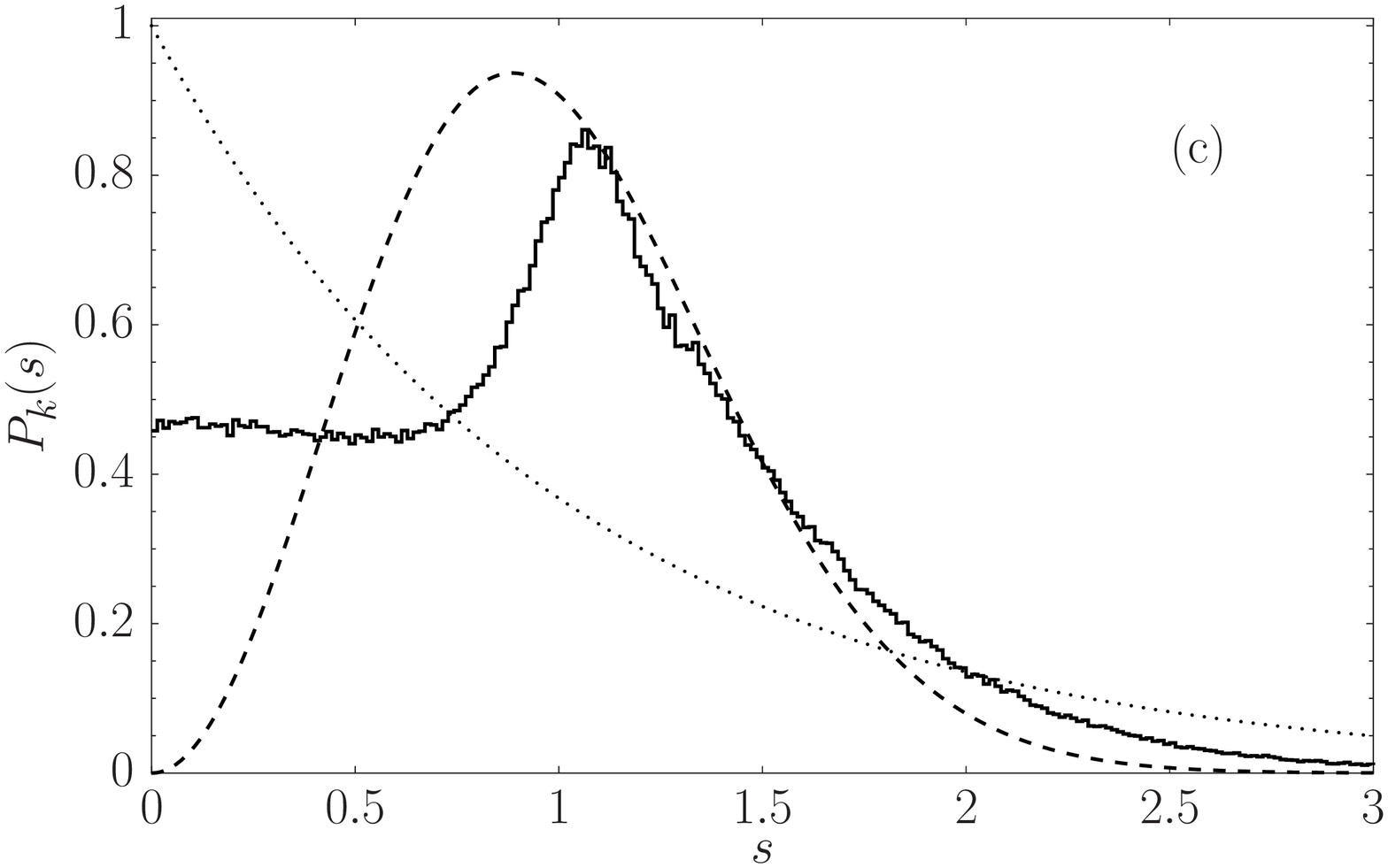}
  \includegraphics[width=8cm]{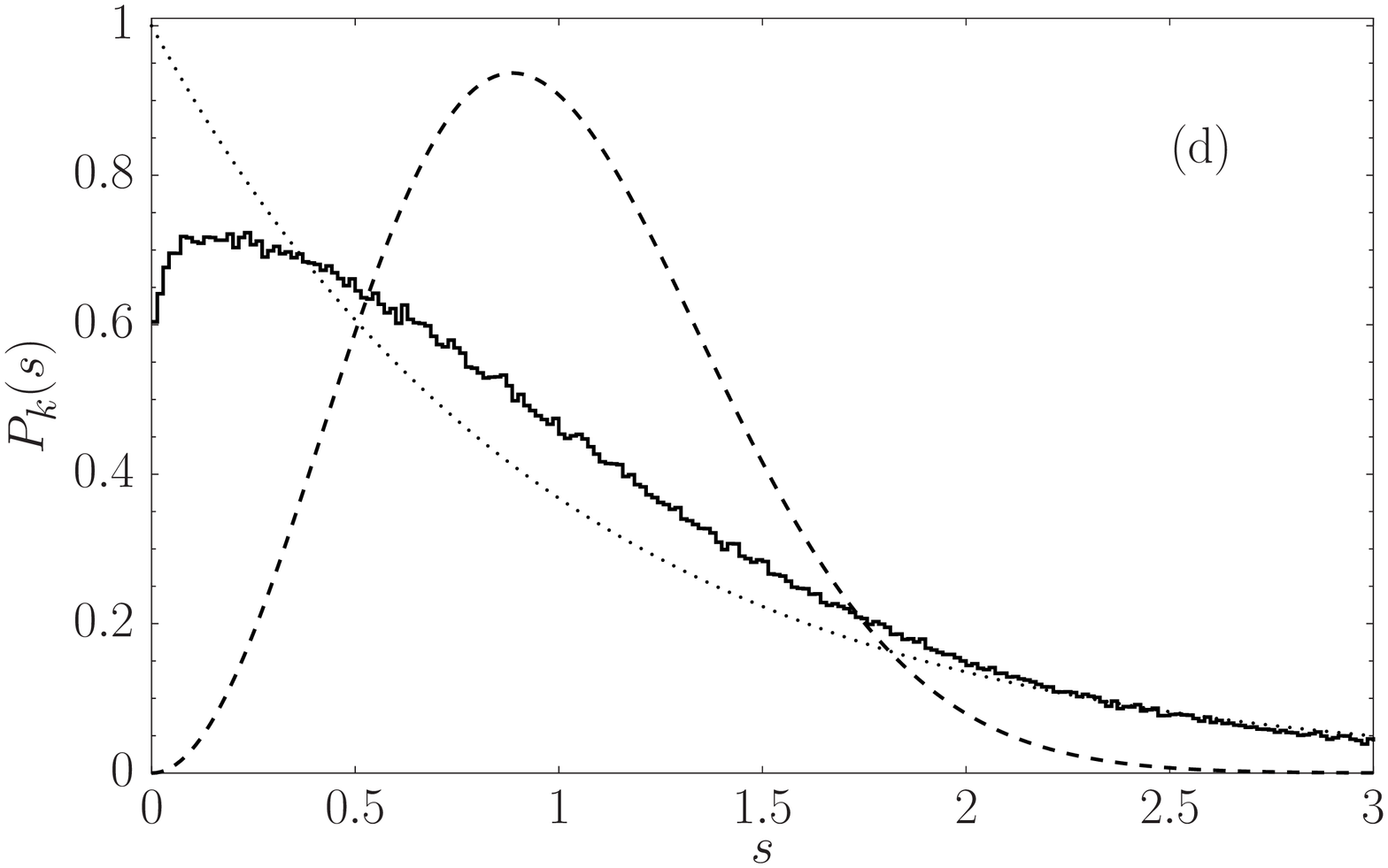}
  \includegraphics[width=8cm]{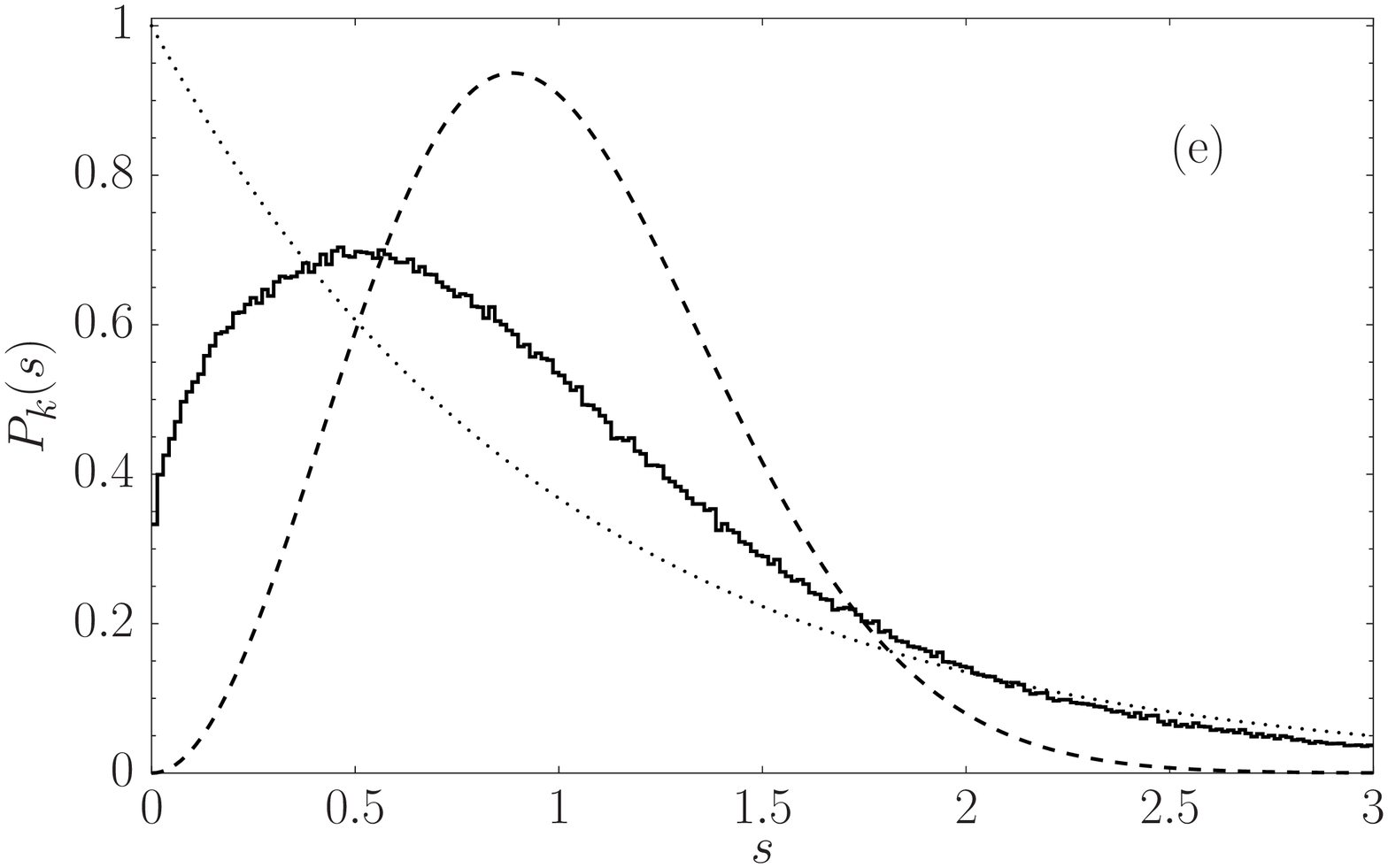}
  \includegraphics[width=8cm]{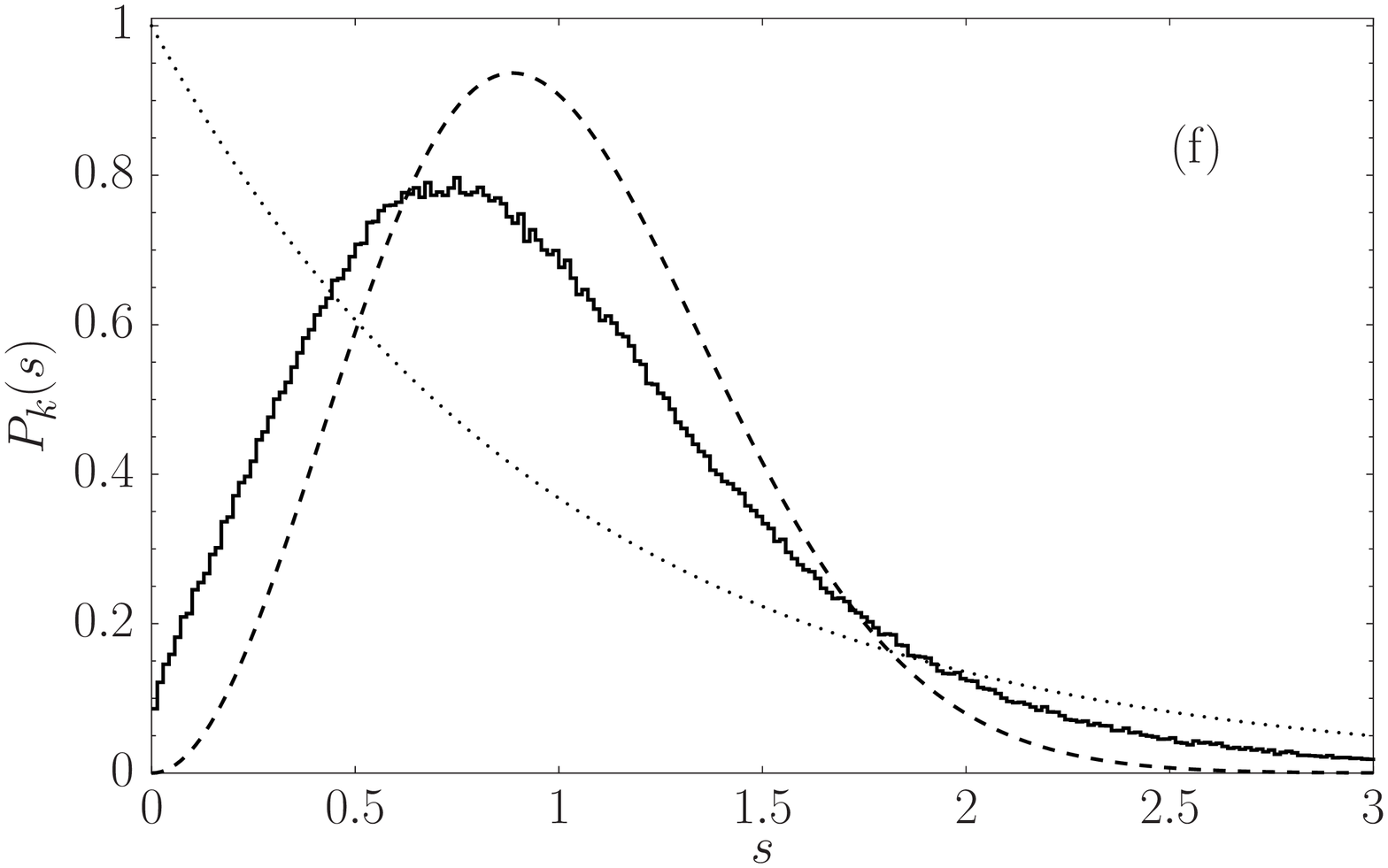}
  \includegraphics[width=8cm]{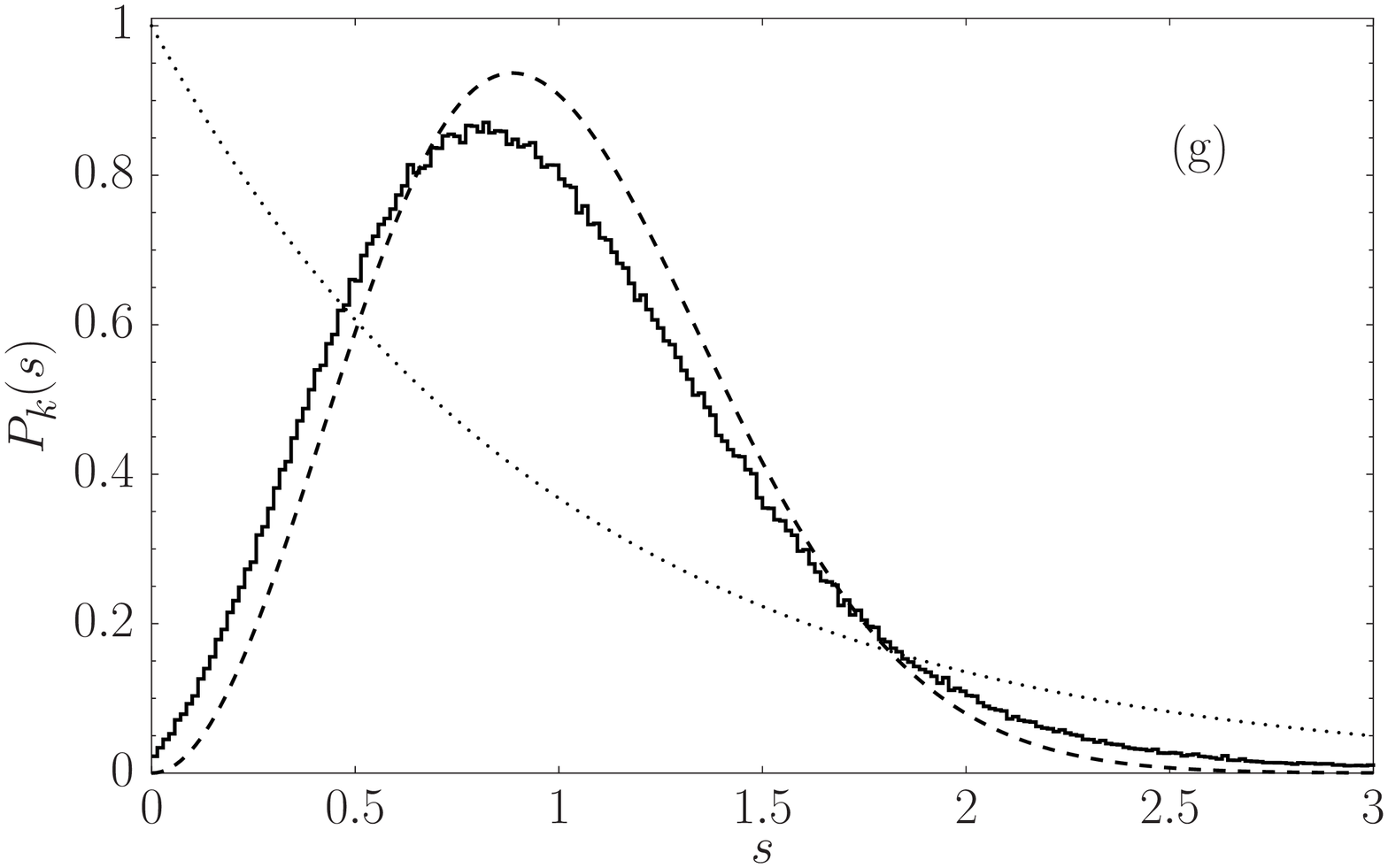}
  \includegraphics[width=8cm]{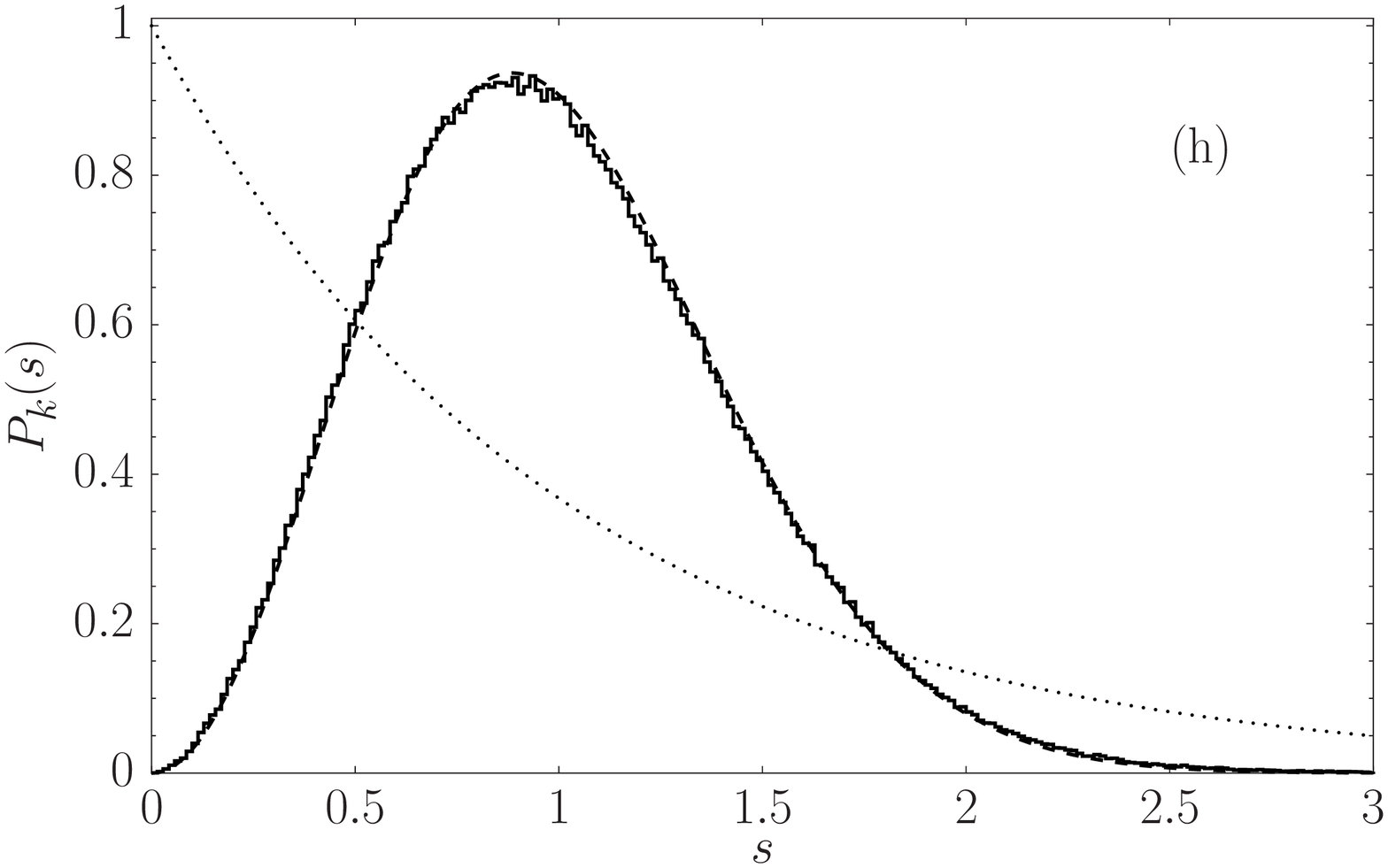}
  \caption{ Same as \Fref{fig1} for $\beta=2$, $n=1000$ and (a)~$k=1$,
    (b)~$k=2$, (c)~$k=10$, (d)~$k=200$, (e)~$k=500$, (f)~$k=700$,
    (g)~$k=800$ and (h)~$k=940$. Notice that the strong peak observed
    in \Fref{fig1} at $s=0$ for $\beta=1$ is absent in this case,
    indicating that its origin is due to time-reversal symmetry. Yet,
    certain degree of level clustering is still observed on a wide
    interval of $k$.}%
  \label{fig2}%
\end{figure*}
 
In Ref.~\cite{Asaga2001}, it was shown that the $k$-body embedded
ensemble of random matrices for bosons is non ergodic in the dense
limit. The dense limit is defined as the limit $n\to\infty$ with $k$
and the number of single-particle levels $l$ fixed. This result was
obtained analytically by considering the fluctuations of the centroids
and variances of individual spectra, which do not vanish in the limit
$n\to\infty$ of infinite Hilbert-space
dimension~\cite{Asaga2001}. Therefore, in the dense limit, ensemble
average and spectral average yield in general different results. The
non-ergodic character of the ensemble in the dense limit is a
consequence of the fact that each member of the ensemble is Liouville
integrable in the classical limit~\cite{BJL2003}.  In this case,
spectral unfolding is the only physically meaningful rectification of
the spectra. In the numerical results described below, we implemented
it by fitting the staircase function of each member of the ensemble
separately with a polynomial of maximum degree $8$.

In~\Fref{fig1}, we present the nearest-neighbor spacing distribution
$P_k(s)$ for various values of $k$, for $\beta=1$ and $n=2000$. These
results were obtained after averaging over $1000$ realizations of
the ensemble. More details can be observed in the accompanying
movie~\cite{Movie1}. In these figures, we have included for comparison
the Poisson distribution and the Wigner surmise for the
GOE~\cite{GMGW1998}.

For $k=1$, the system corresponds to two coupled harmonic oscillators.
Consequently, after unfolding, we obtain the expected distribution for
an equidistant spectrum, i.e., $P_{k=1}(s)=\delta(s-1)$
[\Fref{fig1}a]. As seen in \Fref{fig1}b, for $k=2$, this distribution
changes considerably. It displays a quite large peak at $s=0$, a tail
that decays somewhat slower than the Gaussian tail for larger values
of $s$, and a broad peak around $s=1$ reminiscent of the Dirac delta
obtained for $k=1$. The peak at $s=0$ indicates the occurrence of
quasi-degenerate energy levels and, as we shall demonstrate below, it
is a consequence of the time-reversal symmetry ($\beta=1$) of the
ensemble. Increasing slowly the value of $k$ enhances the level
clustering at $s=0$ and diminishes, shifts, and smoothes the peak at
$s=1$. This is illustrated for $k=10$ in \Fref{fig1}c, where we also
note that the tail of the distribution approaches the exponential
decay characteristic of the Poisson distribution. The local maximum
observed at $s\approx 1$ disappears smoothly by increasing the value
of $k$, being unnoticeable already for $k=75$~\cite{Movie1}.

By increasing the value of $k$, the distribution $P_k(s)$ evolves
smoothly still displaying a strong degree of degeneracy at $s=0$
[cf. Figs.~1d and 1e for $k=200$ and $k=1000$,
respectively]. Eventually, around $k=1150$, a new local maximum of the
distribution is noticeable around $s\approx 0.3$ [\Fref{fig1}f], which
moves toward larger values of $s$ for larger values of $k$; this peak
will become the single maximum of the GOE reached at $k=n$. From here
on, except for the peak at $s=0$, the distribution evolves toward the
GOE results by increasing the value of $k$ (see~\cite{Movie1}),
similarly to the transition observed in the spectral properties of the
system when the dynamics of its classical analog evolves from near
integrable to fully chaotic. Interestingly, the peak at $s=0$ is still
observed for rather large values of $k$. Around $k=1850$
[\Fref{fig1}g] this peak disappears, i.e., level repulsion completely
sets in. Beyond $k=1900$, the distribution corresponds essentially to
that of a GOE. We note that there is no value of $k$ where $P_k(s)$
fully coincides with the Poisson distribution, although it does so for
larger spacings (tail of the distribution) in an extended range of
values of $k$. It is not clear to us how to explain such an
exponential tail for intermediate values of $k$. At the moment, we
believe that this fact may be related with a partial applicability of
the original Berry-Tabor argument, which somehow can not be extended
to all tori (see Ref.~\cite{Marklof2000} for some recent results
discussing the generic aspects of the Berry-Tabor conjecture).

The remarkable property of the nearest-neighbor distributions
described above is the appearance and robustness of the large peak
found around $s=0$. This peak is not only pointing out the lack of
level repulsion, but actually indicating that a relevant part of the
spectrum is degenerate or quasi-degenerate. This peak corresponds to
the prediction of Shnirelman's theorem~\cite{Shnirelman1975}, which
essentially states that smooth-enough time-reversal invariant
($\beta=1$) and integrable Hamiltonian of two degrees of freedom have
an asymptotically multiple spectrum, i.e., quasi-degenerate levels
(see also~\cite{Chi1995}). Note that the assumptions of this theorem
are fulfilled, since each member of the ensemble is Liouville
integrable in the semiclassical limit~\cite{BJL2003}.

\begin{figure}
  \includegraphics[width=8.5cm]{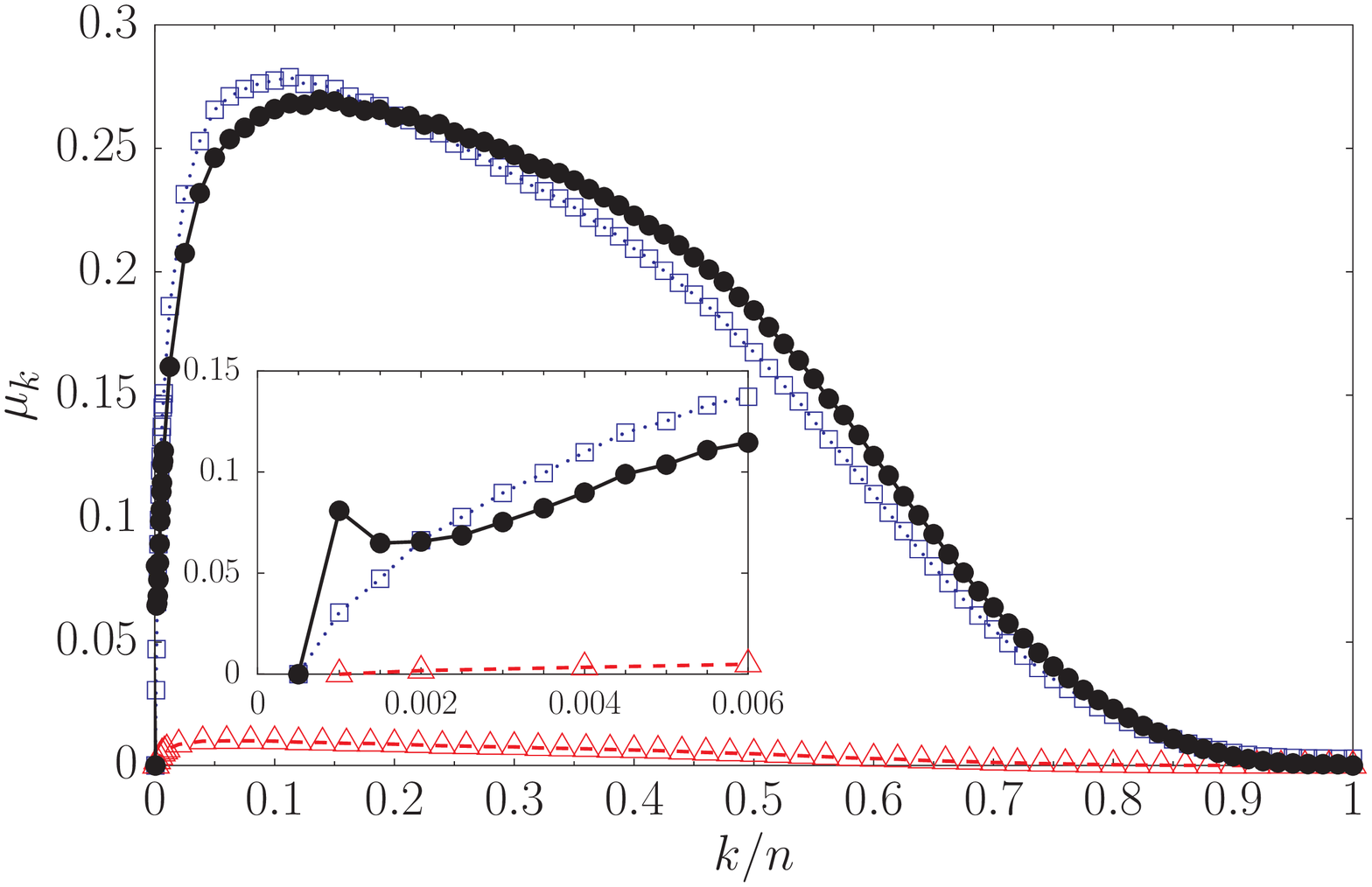}
  \caption{ (Color online) Relative measure $\mu_k$ of the number of
    levels contained within the first four bins of the
    nearest-neighbor spacing distributions as a function of $k/n$. The
    blue curve (dotted curve with squares) corresponds to the
    time-reversal invariant case ($\beta=1$) and the red curve (dashed
    curve with triangles) corresponds to the broken time-reversal case
    ($\beta=2$). The continuous black curve (full circles) represents
    the average number of Shnirelman doublets ($\beta=1$) obtained
    using the symmetry properties of the eigenfunctions. The inset
    shows details for small values of $k$.  }%
  \label{fig3}%
\end{figure}

To completely prove that the peak is indeed Shnirelman's peak, it
suffices to consider the nearest-neighbor distribution $P_k(s)$ for an
ensemble of Hamiltonians ${\hat H_k^{(\beta)}}$ with broken
time-reversal invariance, i.e., $\beta=2$. If time-reversal is
important, the peak should disappear for $\beta=2$. The results are
illustrated in~\Fref{fig2} for different values of $k$, considering
$n=1000$ bosons and $1000$ realizations of the ensemble (see the
corresponding movie~\cite{Movie1} for more details). The figures show
the transition from a picket fence spectrum ($k=1$) to a GUE
($k=n$). In particular, they show the absence of the strong peak at
$s=0$ (Shnirelman's peak), even though as a function of $k$ there is
certain degree of level clustering, which are not quasi-degeneracies
of the type discussed above.  This is further illustrated in
\Fref{fig3}, where we show the relative number of levels $\mu_k$
corresponding to the first four bins of $P_k(s)$ as a function of $k$,
both for $\beta=1$ and $\beta=2$. In this figure we have also included
the average number of degenerate levels, which were identified using
symmetric or antisymmetric combinations of the corresponding
eigenfunctions (cf. Sect.~\ref{eigenf}). Figure~\ref{fig3} implies
that, as a function of $k$, there are different statistical properties
of the degenerate levels. This in turn suggests the use of the
quasi-degenerate levels, i.e., the tunneling splittings, as a possible
measure to test $k$-body interactions in such integrable systems.

\section{Semiclassical limit and the classical phase space}
\label{Sec:PhaseSpace}

\subsection{Semiclassical limit}
\label{semiclassics}

Following Refs.~\cite{Jacob1999,BLS2003}, we write an appropriate
semiclassical limit for the algebraic Hamiltonian ${\hat
  H_k^{(\beta)}}$, which will allow us to identify systematically
time-reversal related symmetric or anti-symmetric combinations of
eigenfunctions. To this end, we symmetrize first ${\hat
  H_k^{(\beta)}}$ with respect to the ordering of the creation and
annihilation operators by exploiting the commutation relations among
the bosonic creation and annihilation operators, typically in the form
$\hat b_r^\dagger \hat b_s = ( \hat b_r^\dagger \hat b_s + \hat b_s
\hat b_r^\dagger-\delta_{r,s})/2$ $(r,s=1,2)$. Then, we use
Heisenberg's semiclassical rules~\cite{Heisenberg1925}
\begin{equation}
\label{eq2}
\hat b_r^\dagger \longrightarrow I_r^{1/2}\exp( i \phi_r) , \qquad
\hat b_r \longrightarrow I_r^{1/2}\exp(-i \phi_r) ,
\end{equation}
where $\phi_r$ is an angle and $I_r$ is its canonically conjugated
momentum. We emphasize the fact that considering the two-level case
($l=2$) implies that the classical associated Hamiltonian has two
degrees of freedom.

The classical Hamiltonian obtained in this way can be written as
${{\cal H}}_k^{(\beta)} (I_1,I_2,\phi_1,\phi_2) = {{\cal
    H}_0}_k^{(\beta)} (I_1,I_2) + {{\cal V}_k}^{(\beta)}
(I_1,I_2,\phi_1,\phi_2)$. Here, ${{\cal H}_0}_k^{(\beta)} (I_1,I_2)$
is a Hamiltonian that depends on the action variables only and is
therefore integrable, and the perturbing term ${{\cal
    V}_k}^{(\beta)}(I_1,I_2,\phi_1,\phi_2)$ carries all the dependence
upon the angles. The first term is associated with all the diagonal
contributions of ${\hat H_k^{(\beta)}}$, while the second one
corresponds to all off-diagonal contributions. These terms are
explicitly given by~\cite{BLS2003}
\begin{equation}
  \label{eq3}
  {{\cal H}_0}_k^{(\beta)} = \sum_{s=0}^k \frac{ v_{s,s}^{(\beta)}}{({\cal N}_s^{(k)})^2}
  {\cal P}_s(I_1-\tfrac{1}{2}, s) {\cal P}_{k-s}(I_2-\tfrac{1}{2}, k-s),
\end{equation}
\begin{eqnarray}
  \label{eq4}
  {\cal V}_k^{(\beta)} & = & \sum_{s>t} 
  \frac{v_{s,t}^{(\beta)} (I_1 I_2)^{(s-t)/2}  }{2 \, {\cal N}_r^{(k)} {\cal N}_s^{(k)}}
  \cos[(s-t) (\phi_2-\phi_1 ) ] \nonumber\\
  & & [{\cal P}_t(I_1-\tfrac{1}{2},s) + {\cal P}_t(I_1-\tfrac{1}{2},t)]\nonumber \\
  & & [{\cal P}_{k-s}(I_2-\tfrac{1}{2},k-s) + {\cal P}_{k-s}(I_2-\tfrac{1}{2}, k-t)].\quad
\end{eqnarray}
In Eqs.~(\ref{eq3}) and~(\ref{eq4}), ${\cal P}_t(I,s)$ are polynomials
of degree $t$ on the variable $I$ defined as
\begin{equation}
  \label{eq5}
  {\cal P}_t(I,s) = \prod_{i=1}^t [I-(s-i)], 
\end{equation}
with $s$ a numerical coefficient satisfying $s\ge t\ge0$. We notice
that the time-reversal symmetry properties of the ensemble are
reflected in the matrix elements $v_{s,t}^{(\beta)}$.

The classical Hamiltonian ${\cal H}_k^{(\beta)}$ is therefore a
general polynomial of degree $k$ on the product of the actions with
random coefficients, modulated by cosine functions whose argument is
$\phi_2-\phi_1$. For $\beta=1$, the matrix elements
$v_{s,t}^{(\beta=1)}$ are real random numbers. Hence, time-reversal
symmetry is manifested through the symmetry under reflection of both
angles, i.e., $\phi_r\to -\phi_r$ for both $r=1, 2$. In the case
$\beta=2$, the matrix $v^{(\beta=2)}$ is complex Hermitian, and the
matrix elements can be written as $v_{s,t}^{(\beta=2)}=
|v_{s,t}^{(\beta=1)}| \exp[i \nu_{r,s}]$, with the random phases
satisfying $\nu_{r,s}=-\nu_{s,r}$ for Hermiticity. Therefore, the
phases $\nu_{r,s}$ for $\beta=2$ can be included into the cosine
functions, manifestly breaking the invariance under simultaneous
reflections.

From Eqs.~(\ref{eq3}) and~(\ref{eq4}), the angle variables appear in
the Hamiltonian ${\cal H}_k^{(\beta)}$ only through the combination
$\phi_2-\phi_1$. In terms of the phase space geometry, this specific
dependence corresponds to one single resonance, which implies the
integrability of the classical Hamiltonian ${\cal
  H}_k^{(\beta)}$. More explicitly, we perform a canonical
transformation to new action and angle variables using the generating
function $W = K \phi_1 +J(\phi_2-\phi_1)$, and obtain $I_1 = K-J$,
$I_2 = J$, $\chi = \phi_1$, and $\psi = \phi_2-\phi_1$. Substituting
these expressions in Eqs.~(\ref{eq3}) and~(\ref{eq4}), the transformed
Hamiltonian depends only on the angle $\psi$. Since the angle $\chi$
does not appear in the transformed Hamiltonian, its canonically conjugated
action $K$ is a conserved quantity, $K=I_1+I_2=n+1$, with $n$ the
number of bosons. Therefore, besides the conservation of the energy
(the Hamiltonian is time independent), we have a second constant of
motion, $K$. It is easy to show that the Poisson bracket between $K$
and ${\cal H}_k^{(\beta)}$ is zero, thus implying that the Hamiltonian
is (Liouville) integrable. For a fixed value of $K$, the reduced
Hamiltonian ${\cal H}_k^{(\beta)}(J,K,\psi)$ is a time-independent
one degree of freedom system with one parameter, which is always
integrable. In the language of symplectic geometry, the reduced
Hamiltonian ${\cal H}_k^{(\beta)}(J,K,\psi)$ is identical to its
normal form. In these variables, the population imbalance, which is a
relevant quantity in the BECs context, is given by
$z=(I_1-I_2)/(I_1+I_2)=1-2J/K$.

\begin{figure*}
  \includegraphics[width=7.8cm]{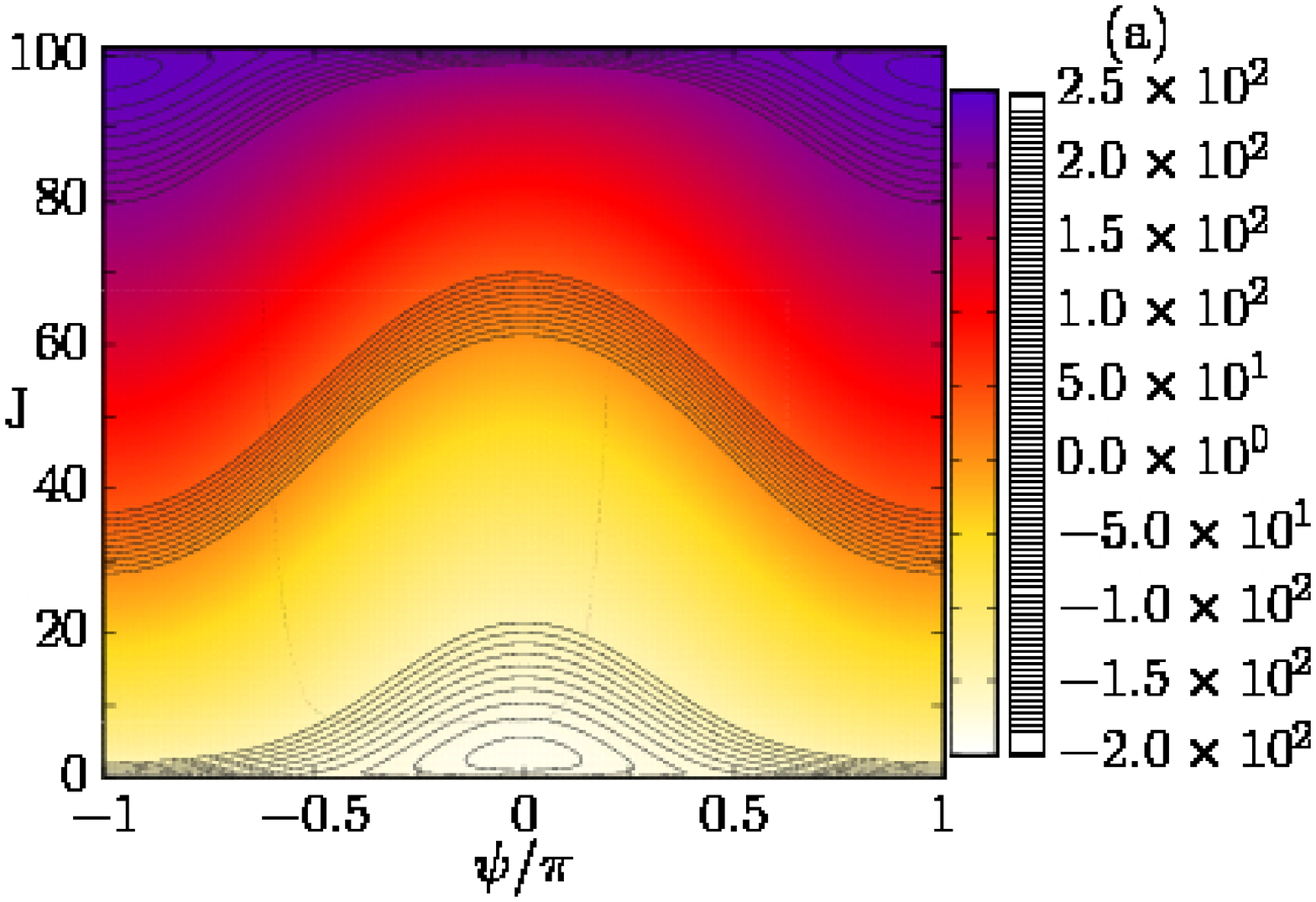}\hspace{0.5cm}
  \includegraphics[width=7.8cm]{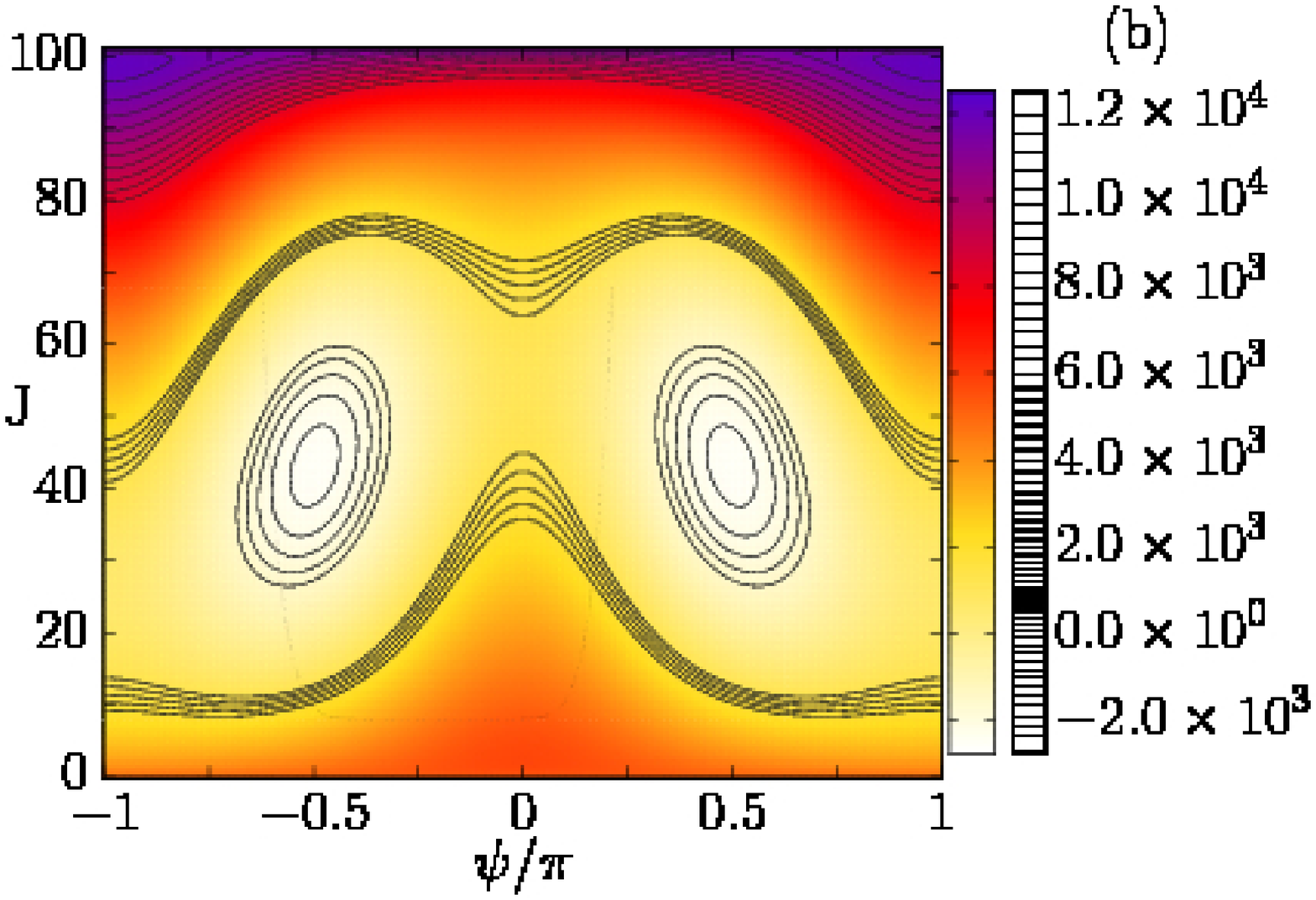}
  \includegraphics[width=7.8cm]{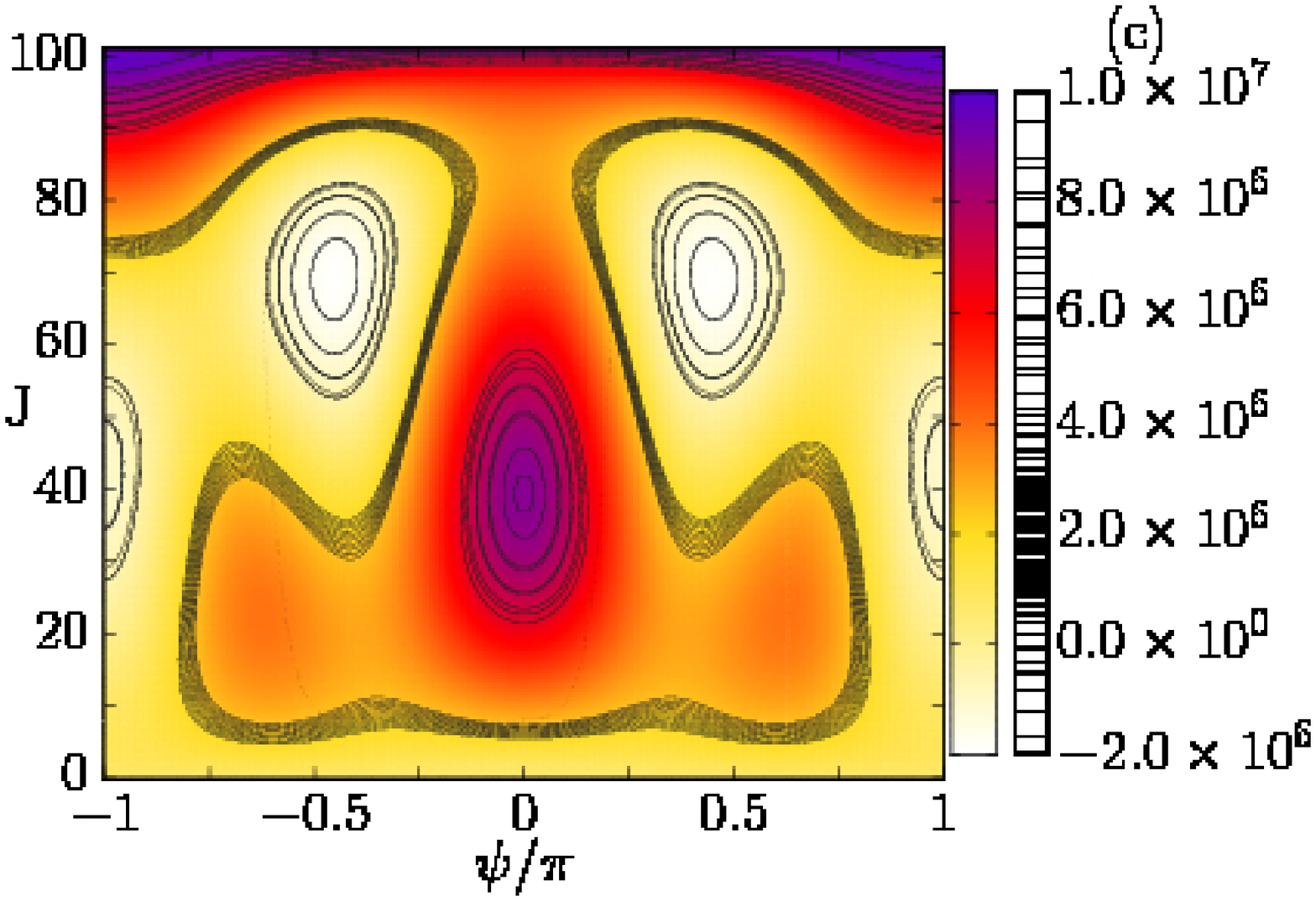}\hspace{0.5cm}
  \includegraphics[width=7.8cm]{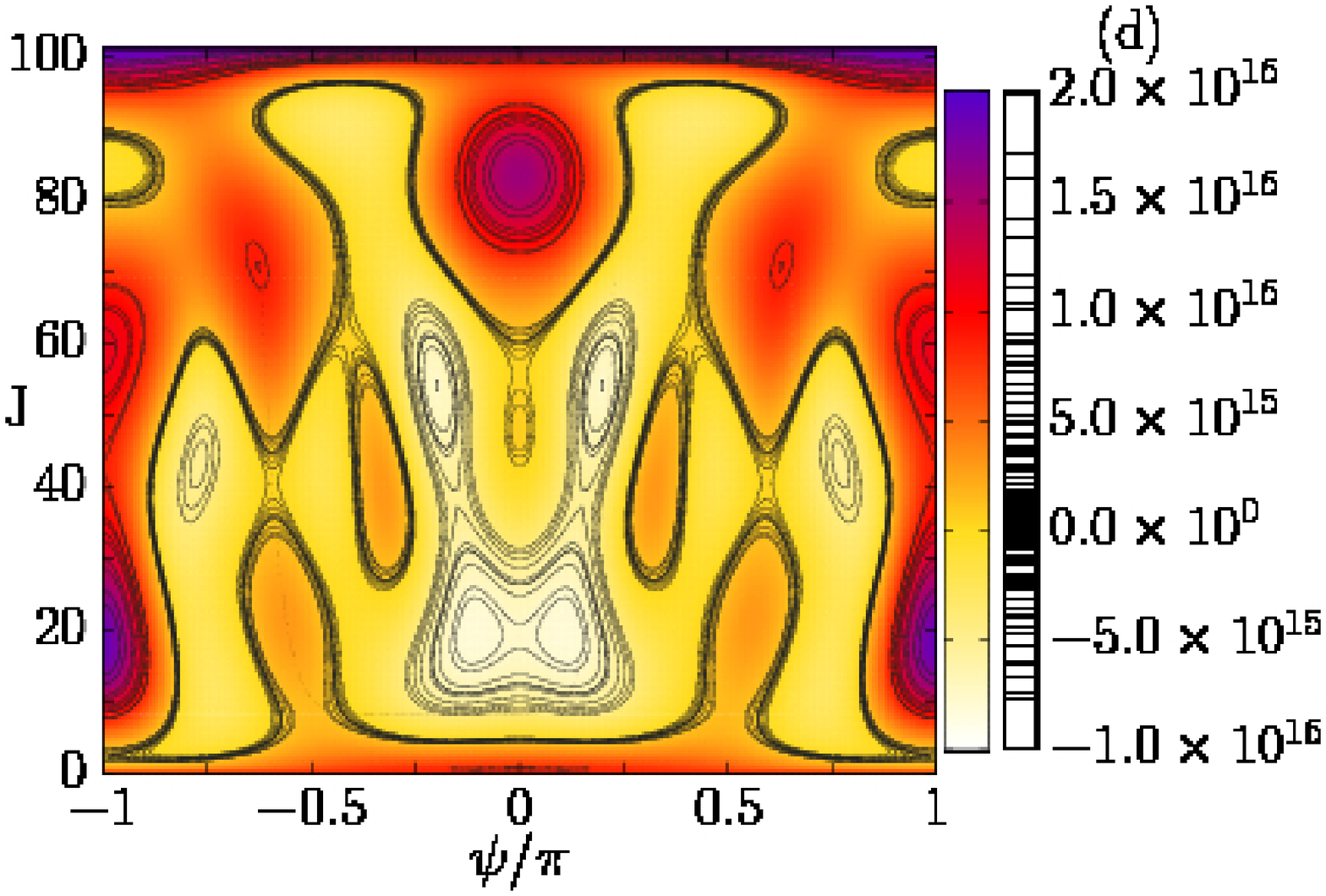}
  \includegraphics[width=7.8cm]{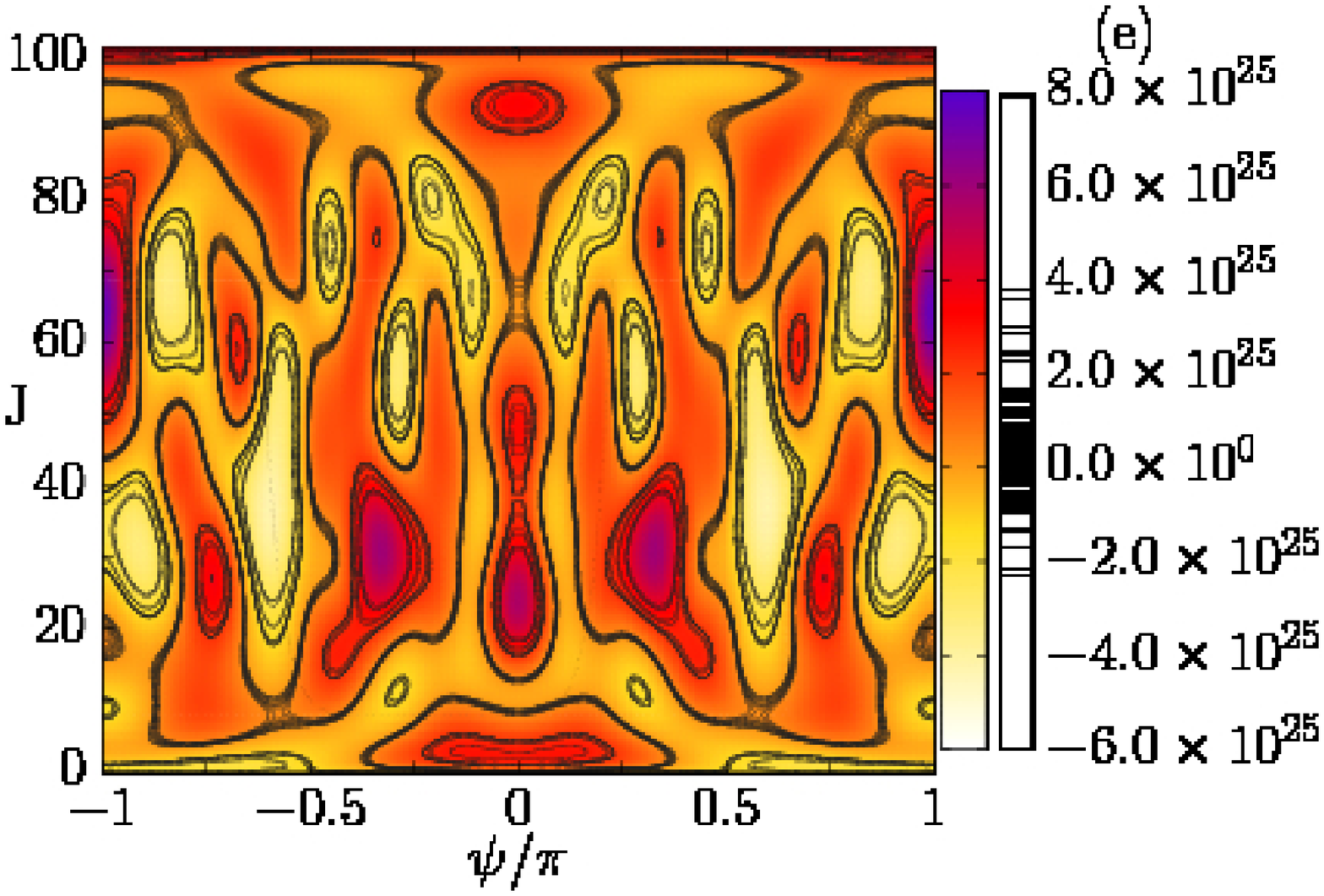}\hspace{0.5cm}
  \includegraphics[width=7.8cm]{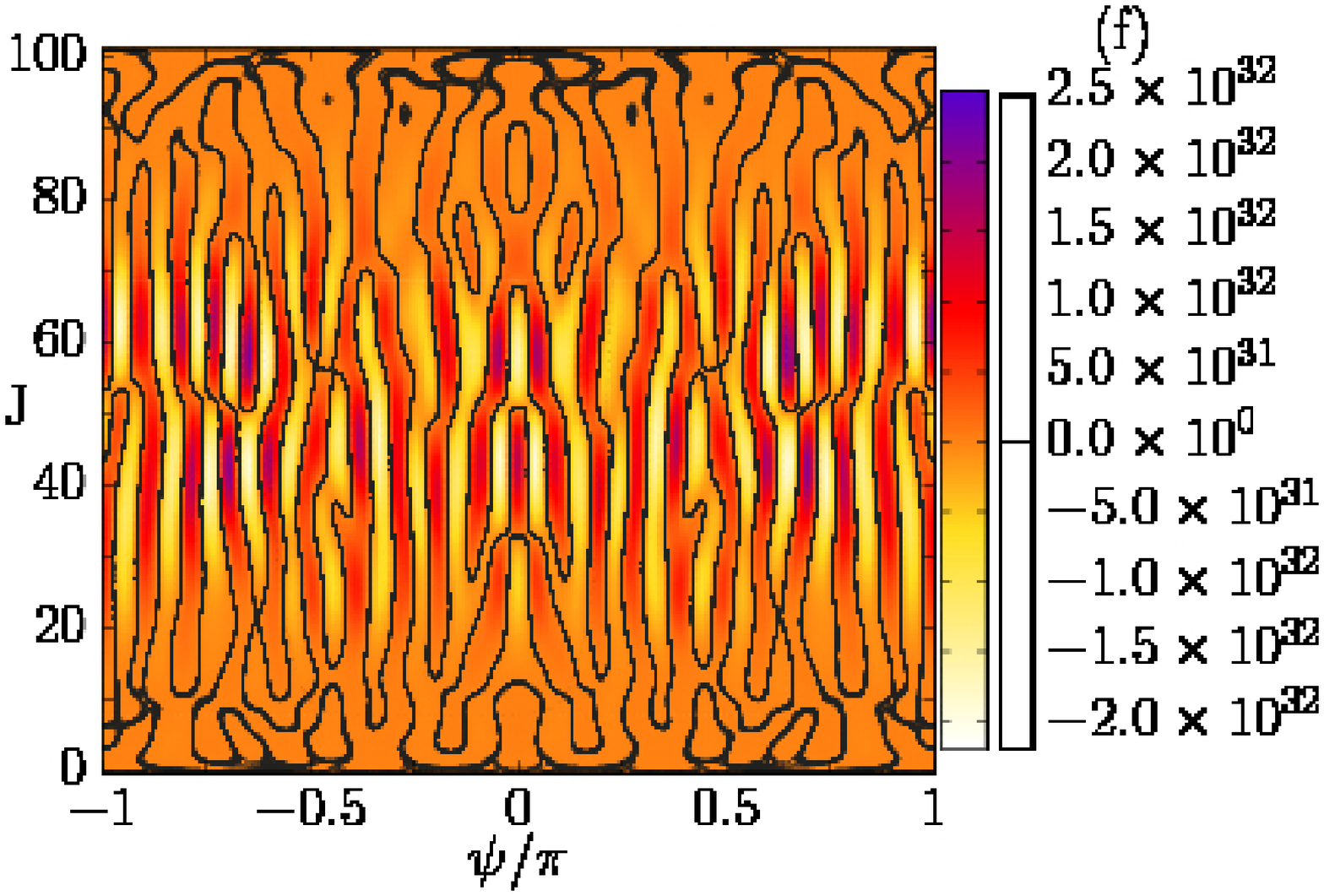}
  \includegraphics[width=7.8cm]{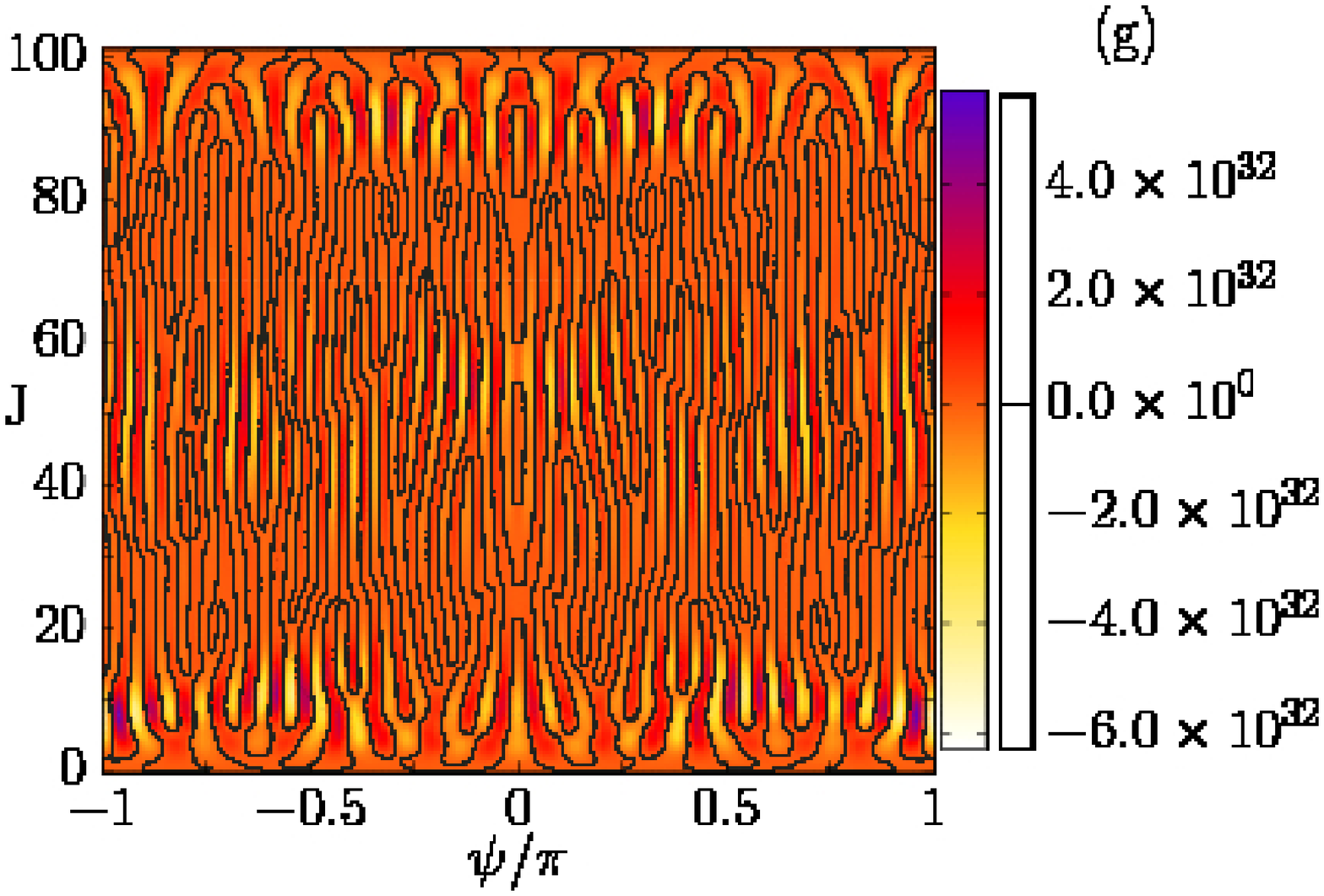}\hspace{0.5cm}
  \includegraphics[width=7.8cm]{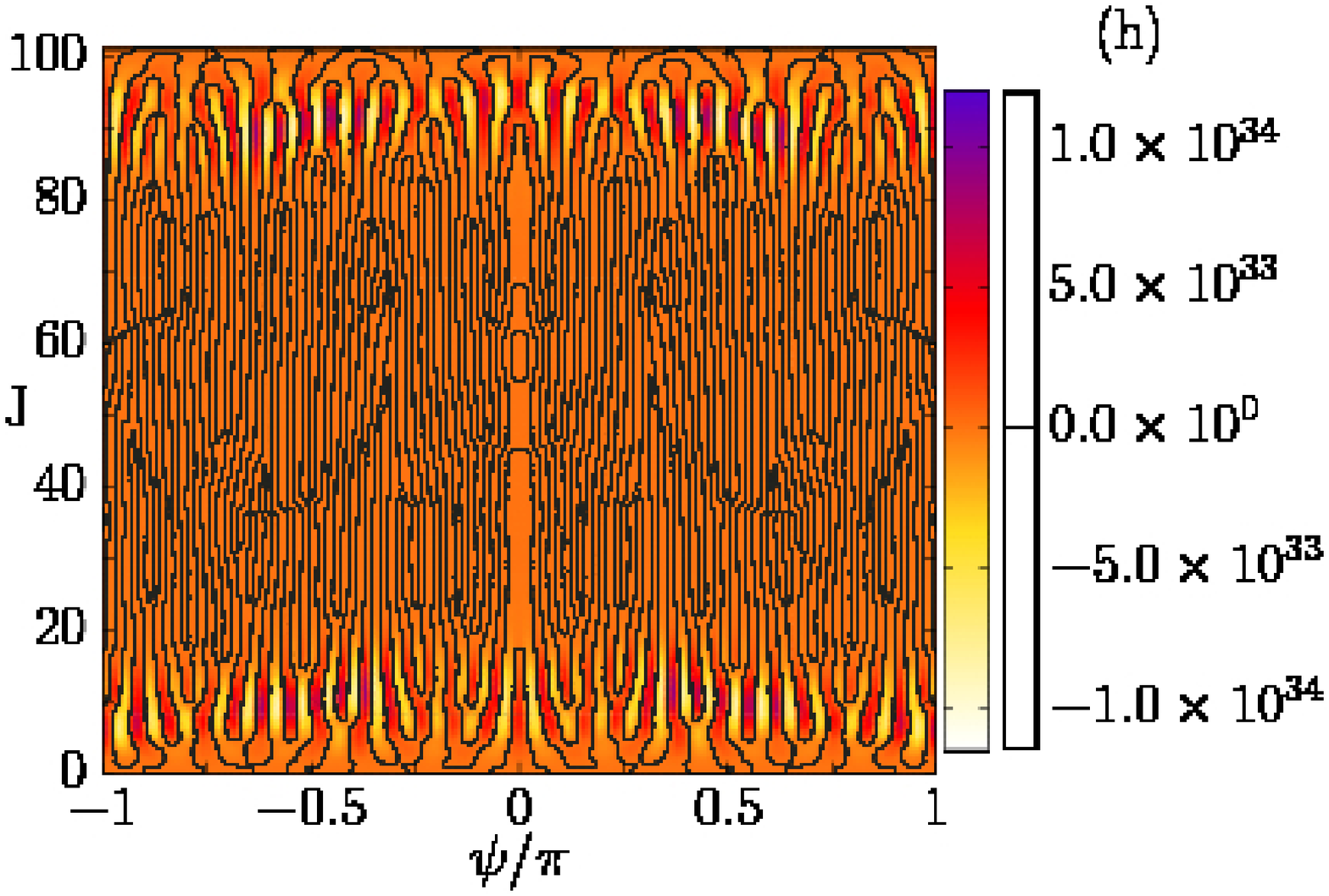}
  \caption{ (Color online) Phase space representation (level curves)
    of the reduced Hamiltonian ${\cal H}_k^{(\beta)}(J,K,\psi)$ for
    $\beta=1$ and $n=100$. (a)~$k=1$, (b)~$k=2$, (c)~$k=4$,
    (d)~$k=13$, (e)~$k=29$, (f)~$k=56$, (g)~$k=81$, (h)~$k=100$. The
    continuous lines represent three groups of 10 levels, taken from
    either edge and from the center of the spectrum.  The color (grey)
    palette displays the overall energy scale of ${\cal
      H}_k^{(\beta=1)}(J,K,\psi)$, while the other one displays the
    full spectrum in the same scale. Note the clear appearance of
    ladders of levels associated with certain organizing centers in
    phase space.  As $k$ is increased, the level curves spread over
    all the available phase space, while the width of the spectrum
    decreases.}%
  \label{fig4}%
\end{figure*}

We finish this discussion mentioning how to relate the action-angle
variables of the reduced system ($J$ and $\psi$) to the actual
coordinates and momenta $q_r$ and $p_r$, $r=1,2$, of the two
single-particle modes. This is done by a lifting
procedure~\cite{Jacob1999}: Integration of the equations of motion of
the reduced system yields $J(t)$ and $\psi(t)$. Undoing the canonical
transformation gives all $I_r(t)$ and $\phi_r(t)$. Then we use the
harmonic expressions
\begin{equation}
\label{eq6}
I_r^{1/2}\exp[\mp i\phi_r]= (q_r \pm i p_r)/\sqrt{2},
\end{equation}
which relate $I_r(t)$ and $\phi_r(t)$ to local coordinates and
momenta, and thus yield the usual representation of the motion of each
mode.

\subsection{Phase space structure in terms of $k$}
\label{phasespace}

The essential features of the classical dynamics can be easily
visualized in Poincar\'e sections of the reduced Hamiltonian. Since
the reduced Hamiltonian is a one-degree of freedom system, this
representation corresponds to the level curves ${\cal
  H}_k^{(\beta)}(J,K,\psi)={\rm const}$. Therefore, the motion of a
given initial condition follows the closed curve that includes the
initial conditions. The explicit appearance of the square-root
factors in \Eref{eq4} makes the classical phase-space bounded, i.e.,
$J \in [0,K]$ and $\psi\in[-\pi,\pi]$, and therefore has the topology
of a sphere. These properties are consistent with the interpretation
of the two-mode Hamiltonian as a spin system.

The phase-space structure of the unperturbed part ${{\cal
    H}_0}_k^{(\beta)}(J,K)$ is trivial since $J$ remains constant.
Therefore, in a (Mercator) representation using the $\psi$-$J$ plane,
the phase space appears foliated in horizontal straight lines for all
$k$, each representing a different level curve. The perturbing term
${\cal V}_k^{(\beta)}(J,K,\psi)$ induces new structure due to the
resonance; the argument of the cosine terms are $k$-dependent integer
multiples of $\psi$. In Figs.~\ref{fig4}, we present the phase-space
structure for various values of $k$ for $\beta=1$ and $n=100$; further
details of the transition are given in the movie~\cite{Movie1}. While
the system is integrable for all $k$ and therefore the phase space is
foliated by invariant tori, the complexity of such tori increases with
$k$.  Figure~\ref{fig4} was constructed choosing a specific (fixed)
random matrix $v^{(\beta)}$ of dimension $n+1$, which defines the case
$k=n$.  For $k=n-1$, we have defined the corresponding $k$-body
interaction by using the same matrix elements $v_{r,s}^{(\beta)}$ for
$r,s \le k$, setting the remaining matrix elements to zero. This
procedure can be iterated to obtain the corresponding matrix $v$ with
any desired value of $k$.

For $k=1$ [\Fref{fig4}a], we have two harmonic-oscillator wells
centered around the two stable fixed points of the system, namely,
around $\psi=0$ for small values of $J$ and $\psi=\pm \pi$ at large
values of $J$. The invariant curves associated to initial conditions
around such wells are self-retracing under time-reversal, i.e., each
one is mapped onto itself under the transformation $\psi\to -\psi$. At
intermediate values of $J$, the level curves are smooth deformations
of the unperturbed invariant curves (straight lines), thus
illustrating Kolmogorov-Arnold-Moser (KAM) theorem. These tori are all
self-retracing under time-reversal.

For $k=2$ [\Fref{fig4}b], the two harmonic fixed points (at the poles
of the phase-space sphere) are still observed. Yet, at intermediate
values of $J$, we notice two other fixed points close to $\pm \pi/2$
and the associated KAM-tori surrounding them. These structures are
non-self-retracing under time reversal. Consequently, those tori that
satisfy the Einstein-Brillouin-Keller (EBK) quantization rule
\begin{equation}
\label{eq7}
S(E_i) = \frac{1}{2\pi} \oint J(E_i) {\rm d}\psi = \frac{1}{2\pi}
(\kappa_i+\frac{\alpha_i}{4}),
\end{equation}
where $\kappa_i$ is an integer and $\alpha_i$ is the associated Maslov
index~\cite{Littlejohn87}, yield two degenerate ladders of levels,
each pair associated with a torus and its associated time-reversed
partner; quantum effects lift the degeneracy and produce the splitting
of the quasi-degenerate levels. These doublets are precisely
Shnirelman doublets. In \Fref{fig4}c, we present the case $k=4$, which
displays the appearance of new non-self-retracing wells which yield
ladders of double degenerate levels. Note that there are also
self-retracing tori. The ladders of levels associated with these may
display accidental degeneracies with the levels of other
ladders. These accidental degeneracies do not correspond to Shnirelman
doublets since they may exist for the case of broken time-reversal
invariance; this explains the level clustering observed for $\beta=2$
(see \Fref{fig3}).

Increasing further the value of $k$ increases the complexity of phase
space~\cite{Movie1}. New self-retracing and non-self-retracing wells
appear in phase space, leading to Shnirelman doublets and perhaps
accidental degenerate levels, respectively [see \Fref{fig4}d]. Around
$k\gtrsim 20$, the number of stable fixed points begins to increase
more rapidly, namely, quadratically with respect to $k$. Increasing
$k$ seemingly leads to a clustering in the {\it sub-tropical region}
around the equator ($J\approx K/2$) of the majority of the stable
fixed points (see \Fref{fig4}f for $k=56$).  Eventually, around
$k\approx 70$, the stable fixed points begin to migrate to the polar
regions of the phase-space sphere, with essentially all of them in
that region for $k\ge 81$ [see Figs.~\ref{fig4}g and \ref{fig4}h].

\begin{figure}
  \includegraphics[width=8.5cm]{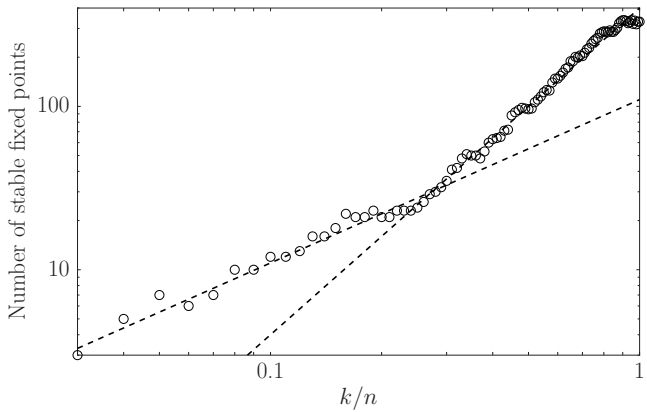}
  \caption{Log-log plot showing the growth of the number of stable
    fixed points in terms of $k$ for $n=100$. The straight lines
    included have slope equal to 1 or 2. Around $k\gtrsim 20$ the
    initial linear growth rate becomes quadratic. }%
  \label{fig5}%
\end{figure}

As illustrated in Figs.~\ref{fig4}, the width of the spectrum is not
constant with respect to $k$~\cite{Asaga2001}. For small $k$, the
width is of the order of the energy scale spanned by ${\cal
  H}_k^{(\beta)}(J,K,\psi)$ [see the {\it spectral palette} of
Figs.~\ref{fig4}a-\ref{fig4}d]; this allows to visually identify
different regions of the phase space where the eigenvalues are
located. However, for larger values of $k$, the width is much smaller
in comparison to the full classical energy interval. In fact, for
$k=n$ [\Fref{fig4}h], the width is proportional to
$n^{1/2}$~\cite{GMGW1998}, while the full energy range is several
orders of magnitude larger. Indeed, Eqs.~(\ref{eq3}) and~(\ref{eq4})
involve homogeneous polynomials of degree $k$ on $J$ and $J$ spans
the interval $[0,n+1]$, thus spanning a huge energy interval. In turn,
violent oscillations are due to the angular dependence, which is
linear on $k$. Note that for $k=n$ [\Fref{fig4}h], the level curves
associated with the energies of the spectrum spread essentially over
all the phase space in such a way that it is not possible to
distinguish one from another anymore, independently of their position
in the spectrum.

The growth rate of the number of stable fixed points is illustrated in
a log-log plot in \Fref{fig5}. The plot displays an initial linear
growth of the number of stable islands, which beyond $k\approx 20$
becomes quadratic in $k$. This figure provides a solid base for the
heuristic arguments of Ref.~\cite{BLS2003}. There, it was argued that
an expected quadratic growth in $k$ of the number of stable fixed
points (now shown in \Fref{fig5}) diminishes the available phase-space
area around the center of the wells. This implies that the EBK states
which originally were found around those wells, will now be defined on
tori which are spread over more extended regions in phase space.
Notice that this argument also shows that it is more difficult to have
quantized states associated with non-self-retracing orbits for large
values of $k$ and thus explains that beyond certain $k$, the number of
Shnirelman pairs diminishes fast and eventually vanishes.

\section{Shnirelman doublets and the statistics of their splittings}
\label{Sec:Splittings}

As shown above, for $\beta=1$ Shnirelman doublets appear and
correspond to the quantization of non-self-retracing periodic orbits,
i.e., orbits which are not mapped onto their selves under
time-reversal invariance ($\psi\to -\psi$). Yet, these are not the
only quasi-degenerate levels found since there are also accidental
degeneracies involving two distinct self-retracing tori that just
happen to have the same energy. Note that Shnirelman doublets
disappear for $\beta=2$, while the accidental degeneracies
persist. Therefore, in order to distinguish the true Shnirelman
doublets, we must consider the corresponding eigenfunctions, in a
representation where the time-reversal invariance is appropriately
manifested. For this purpose, we first analyze the structure of the
eigenfunctions of the quasi-degenerate states using a plane-wave
decomposition which is straightforward to interpret in semiclassical
terms~\cite{Jacob1999}.  Once we have classified the
quasi-degeneracies, we address the question of the statistics of the
Shnirelman splittings.

\subsection{Plane-wave decomposition of the wave functions and
  time-reversal symmetry}
\label{eigenf}

The Hamiltonian \Eref{eq1} is conveniently expressed in the number
occupation basis or Fock basis. For a given number of bosons $n$, we
denote by $|n_1, n_2\rangle$ the state having $n_1$ bosons in the
first single-particle state and $n_2$ bosons in the second and
$n=n1+n2$. Upon diagonalization, the eigenfunctions of each
realization of the ensemble are written as linear combinations of
these basis states and have the form $ |\Phi_r\rangle =
\sum_{n_1+n_2=n} c_{n1,n2}^r |n_1,n_2\rangle$.

The idea now is to use a representation where the symmetry properties
of the time-reversal invariance are manifested. To this end we recall
that semiclassically the number states can be represented as plane
waves on the configuration torus (defined by the angle variables),
namely, $|n_i\rangle\to \exp (i n_i \phi_i) |\phi_1,
\phi_2\rangle$~\cite{Jacob1999}. Hence, the eigenstates can be written
as $| \Phi_r \rangle = \sum_{n_1+n_2=n} c_{n1,n2}^r \exp[ i (n_1
\phi_1 +n_2 \phi_2)] |\phi_1, \phi_2\rangle$. Since the total boson
number is conserved, the associated dimensional reduction is
implemented with the same canonical transformation described for the
classical action-angle variables, which is a point
transformation. Then, the eigenfunction of the $r$th excited state is
written as%
\begin{equation}
  \label{eq8}
  \Phi_r(\psi) = \exp( i n \chi) \sum_{n_2} c_{n-n_2,n2}^r \exp ( i n_2 \psi) .
\end{equation}
In \Eref{eq8}, the factor $\exp(i n \chi)$ is a common phase factor
for all eigenstates, which can therefore be ignored, implying that the
wave functions are functions only of the angle $\psi$. Equation
(\ref{eq8}) defines the reduced representation of $r$th wave
function.

\begin{figure}
  \includegraphics[width=8.0cm,height=18cm]{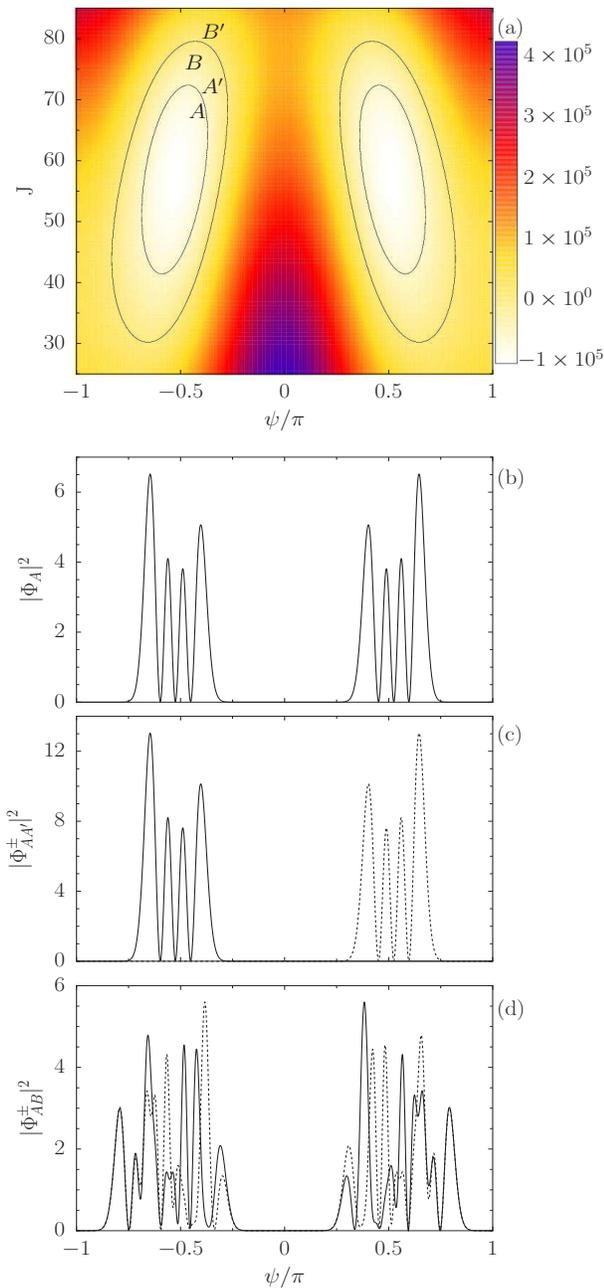}
  \caption{ (Color online) (a)~Classical phase space representation
    for $k=3$ showing two pairs of non-self-retracing tori. The labels
    indicate the symmetry-related eigenfunctions. The color (grey)
    code is related to the full classical energy-scale of this case.
    (b)~Reduced representation of the square modulus of the
    eigenfunction $A$. (c)~Linear combinations~(\ref{eq9})
    and~(\ref{eq10}) with respect to levels $A$ and $A'$; the figure
    shows that the linear combinations lie on opposite sides of the
    $\psi$ axis and therefore the eigenfunctions are a Shnirelman
    doublet. (d)~Result of the linear combinations when considering
    two non-symmetry related nearby levels $A$ and $B$.  }%
  \label{fig6}%
\end{figure}
 
As discussed above, Shnirelman doublets are related to
non-self-retracing tori (under the transformation $\psi\to-\psi$)
that sustain a state. The corresponding eigenfunctions $\Phi_r(\psi)$
and $\Phi_{r^\prime}(\psi)$ appear as mixtures of wave functions
localized around each symmetry-related torus. Therefore, we consider
the linear combinations
\begin{eqnarray}
  \label{eq9}
  \Phi_{r,r^\prime}^{+}(\psi) & = & \frac{1}{\sqrt{2}}{\rm Re} 
  (\Phi_{r}(\psi) + i \Phi_{r^\prime}(\psi))\\
  \label{eq10}
  \Phi_{r,r^\prime}^{-}(\psi) & = & \frac{1}{\sqrt{2}} {\rm Re} 
  (\Phi_{r}(\psi) - i \Phi_{r^\prime}(\psi)).
\end{eqnarray}

In order to identify Shnirelman doublets we proceed as follows: First,
we identify energy levels which lie very close together, which in
practical terms means within the first few bins of the
nearest-neighbor distribution measured in units of the mean-level
spacing. One would naively think that Shnirelman doublets appear as
consecutive levels; yet, accidental degeneracies due to other tori may
have energies in between those of the doublets. This happens rather
frequently for $k/n\gtrsim 0.2$, where the number of stable fixed
points grows quadratically on $k$.  Therefore, we must check not only
degeneracy with respect to the nearest level, but within a wider
range. Then, for each candidate $r$ of a Shnirelman doublet, we
consider a second level $r^\prime$ and construct the linear
combinations given by Eqs.~(\ref{eq9}) and~(\ref{eq10}). If and only
if the functions $\Phi_{r,r^\prime}^{+}(\psi)$ and
$\Phi_{r,r^\prime}^{-}(\psi)$ are concentrated on one side of the
$\psi$ axis (either positive or negative), and among them they are in
opposite sides, then we say that the levels correspond to a Shnirelman
doublet. In this case, the functions $\Phi_{r,r^\prime}^{+}(\psi)$ and
$\Phi_{r,r^\prime}^{-}(\psi)$ are said to be related by the
time-reversal transformation $\psi\to -\psi$.

This method is illustrated in \Fref{fig6}. In \Fref{fig6}(a), we plot
the classical phase-space representation of the non-self-retracing
tori corresponding to two pairs of Shnirelman doublets belonging to
the same ladder. Figures~\ref{fig6}(b)-\ref{fig6}(d) display the reduced
representation of the modulus square of some linear combinations
involving these states. Figure~\ref{fig6}(b) displays one of the states
corresponding to the tori $A$. In \Fref{fig6}(c), we present the linear
combinations (\ref{eq9}) and~(\ref{eq10}) involving the states $A$ and
$A'$; the resulting states are localized on either side of the
$\psi=0$ line, from where it is clear that these states are related by
time reversal invariance. Finally, in \Fref{fig4}(d), we display the
linear combinations involving two states $A$ and $B$ which belong to
the same ladder but are not related by time-reversal invariance.

\subsection{Statistical properties of Shnirelman splittings}
\label{splitt}
\begin{figure*}
  \includegraphics[width=8cm]{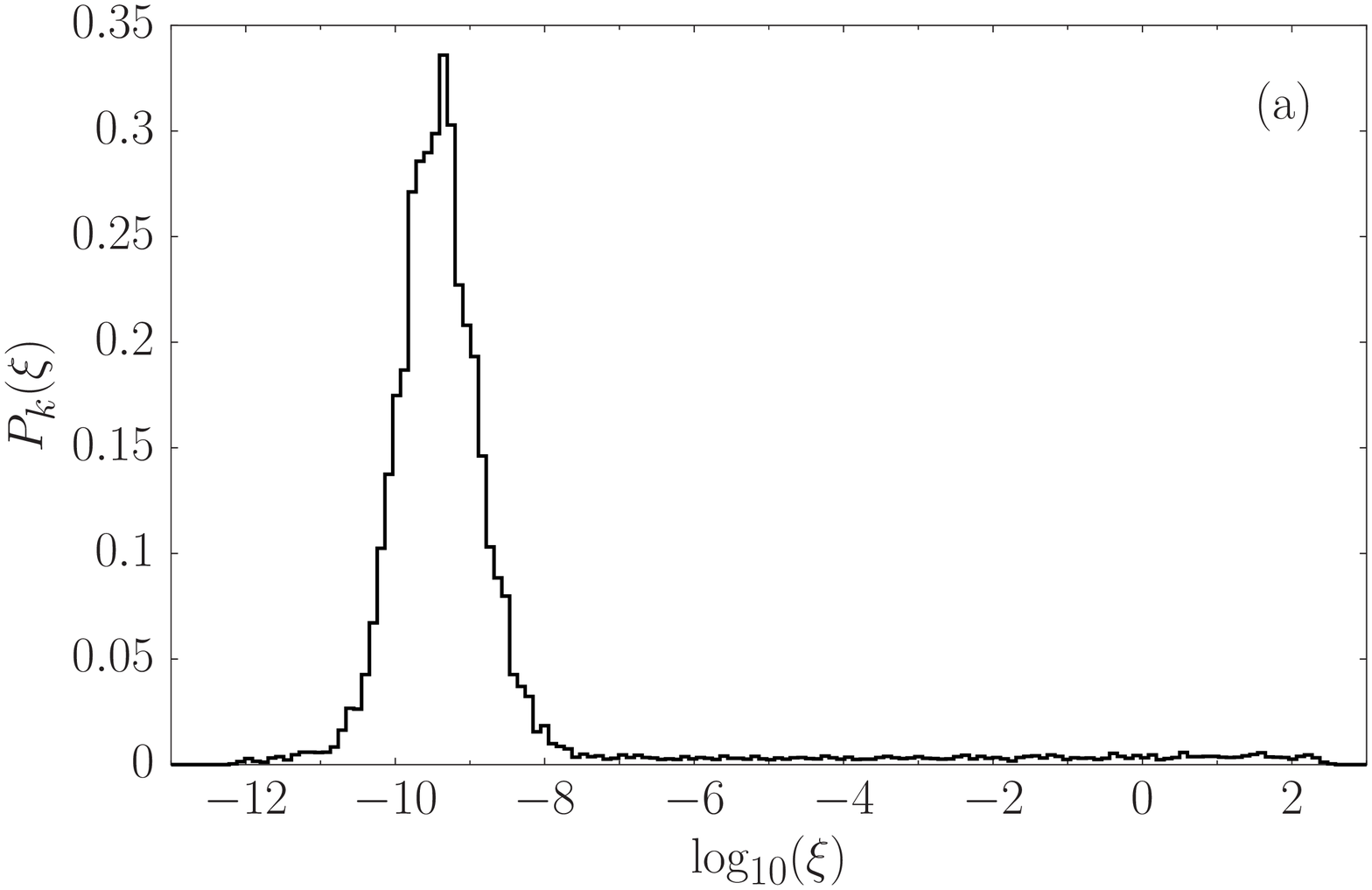}
  \includegraphics[width=8cm]{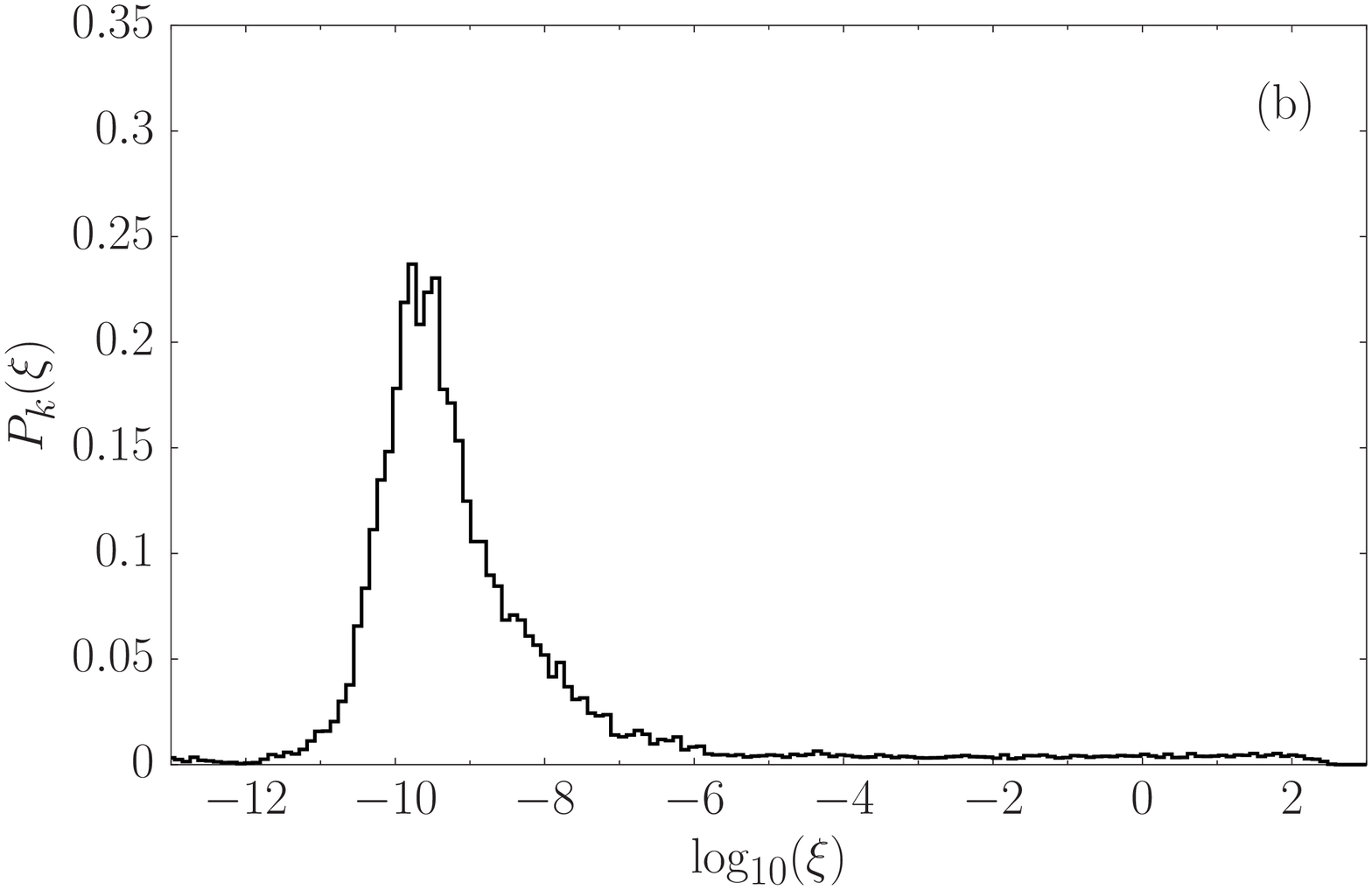}
  \includegraphics[width=8cm]{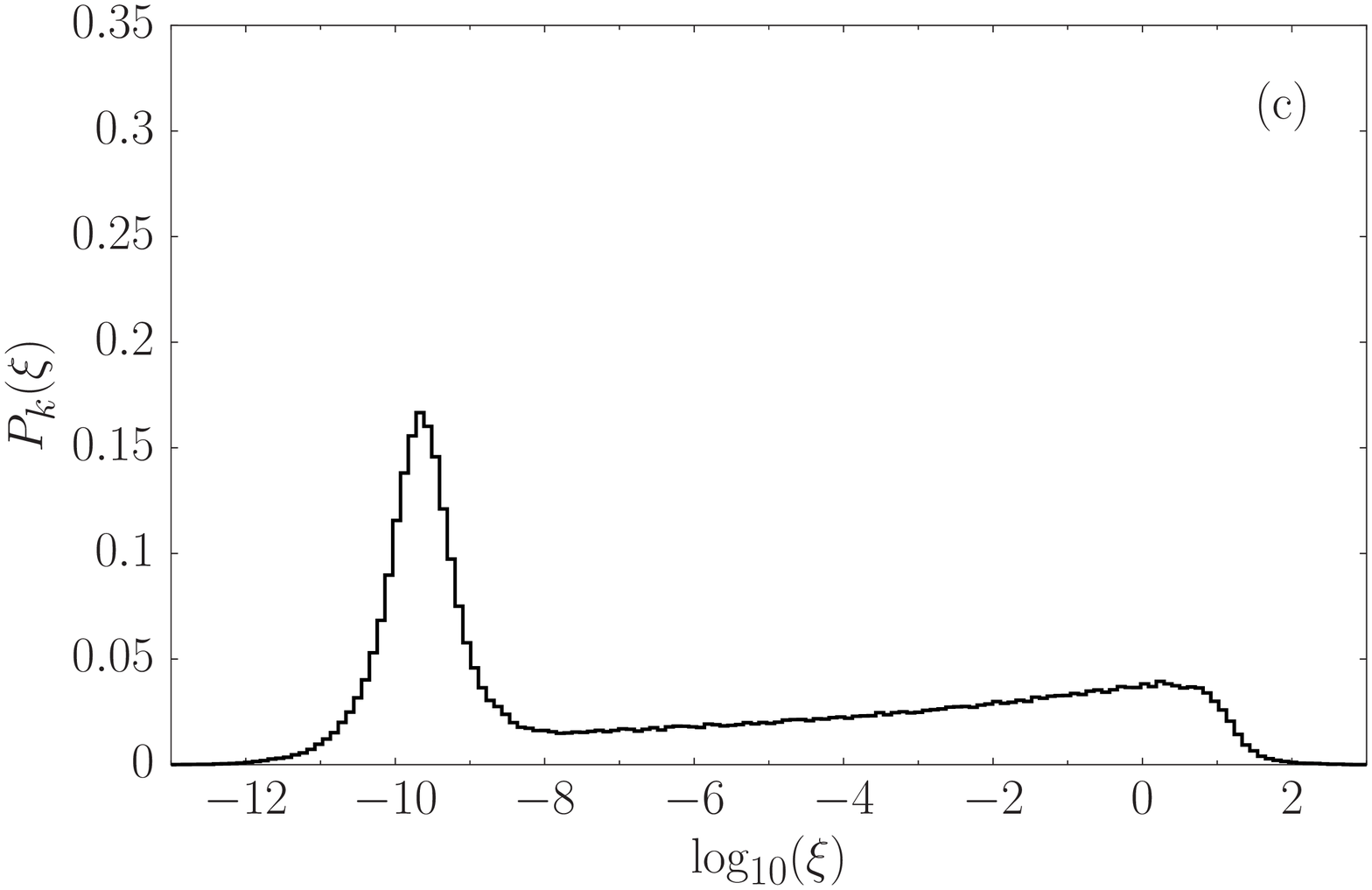}
  \includegraphics[width=8cm]{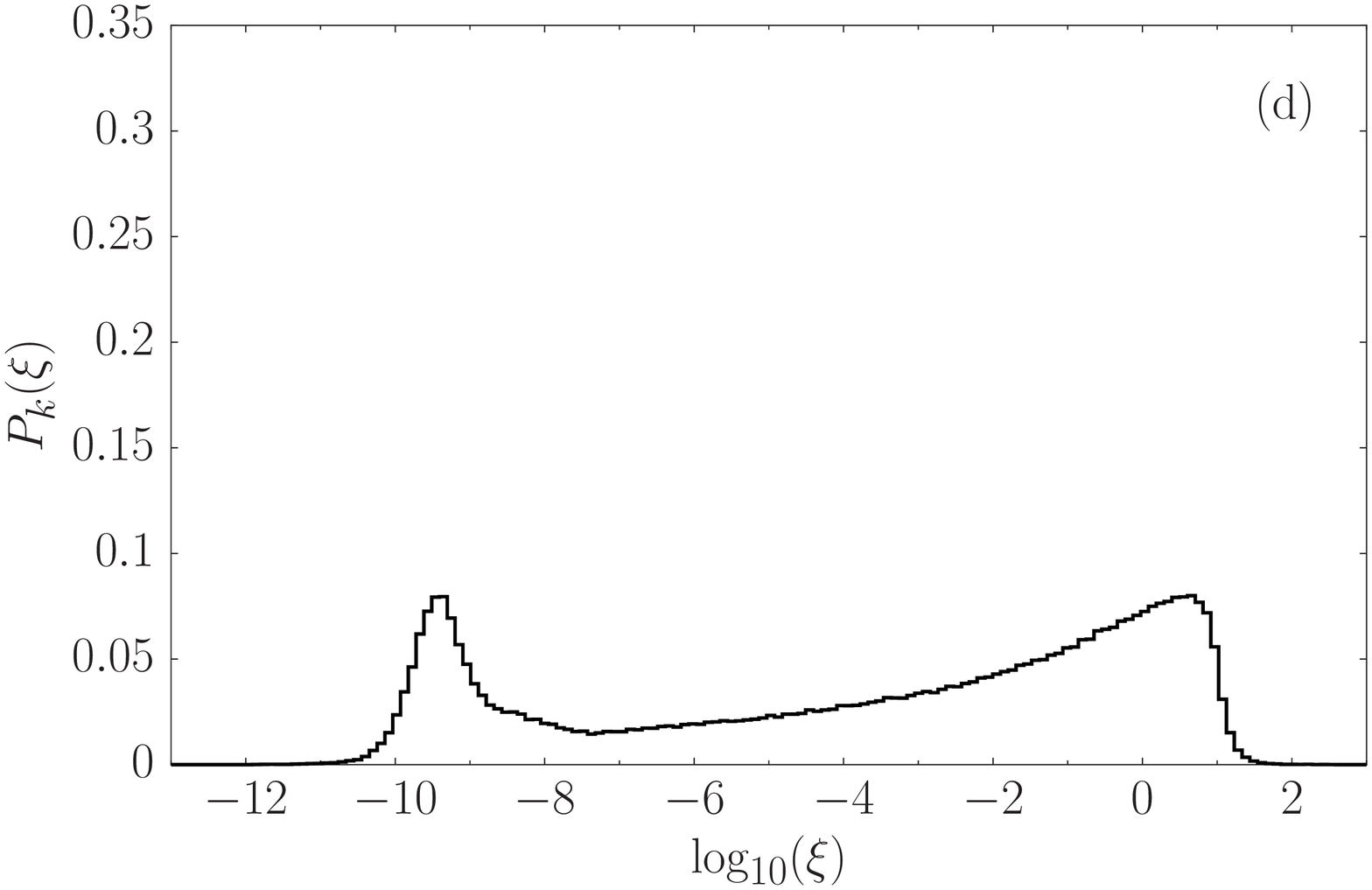}
  \includegraphics[width=8cm]{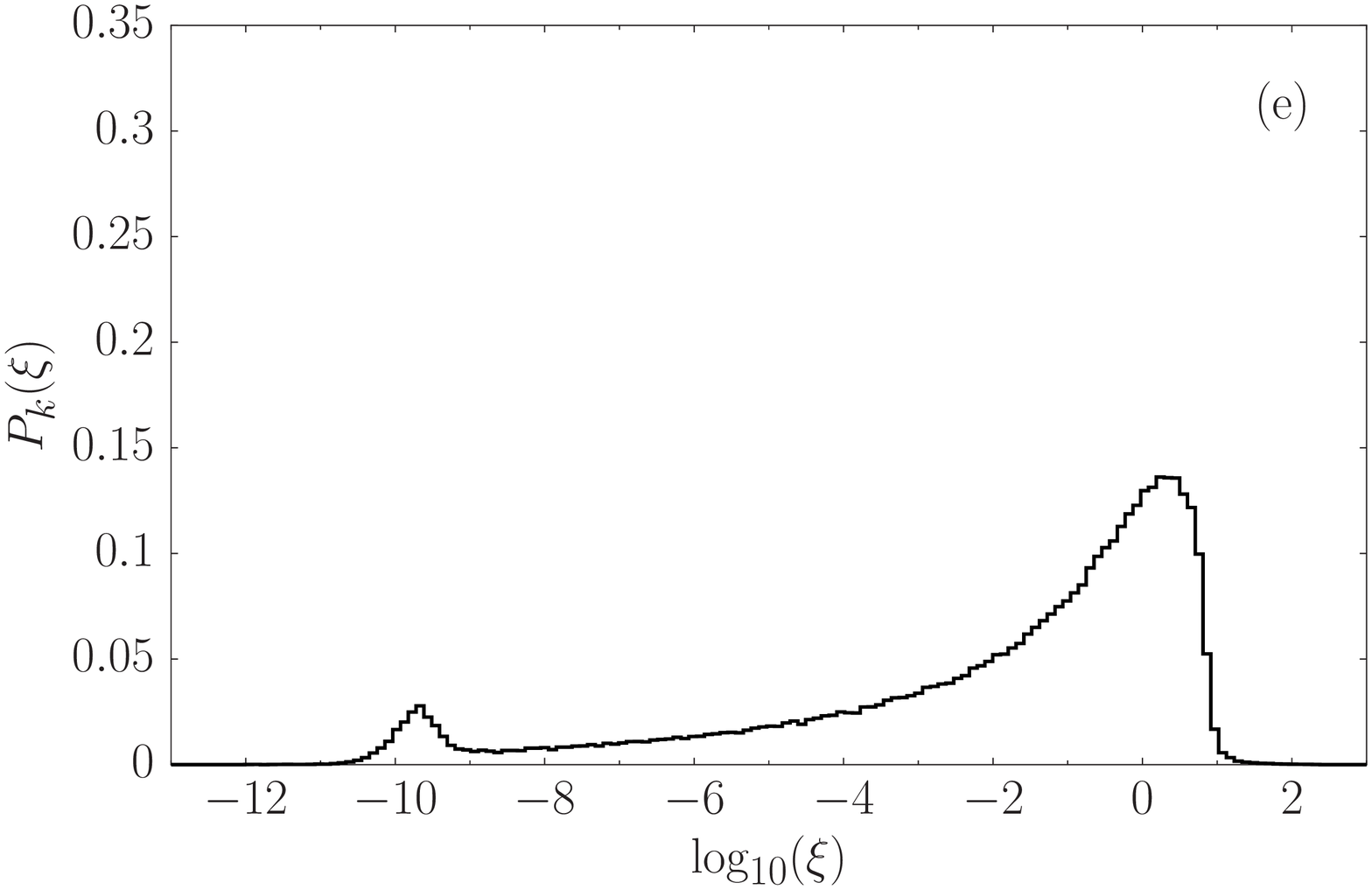}
  \includegraphics[width=8cm]{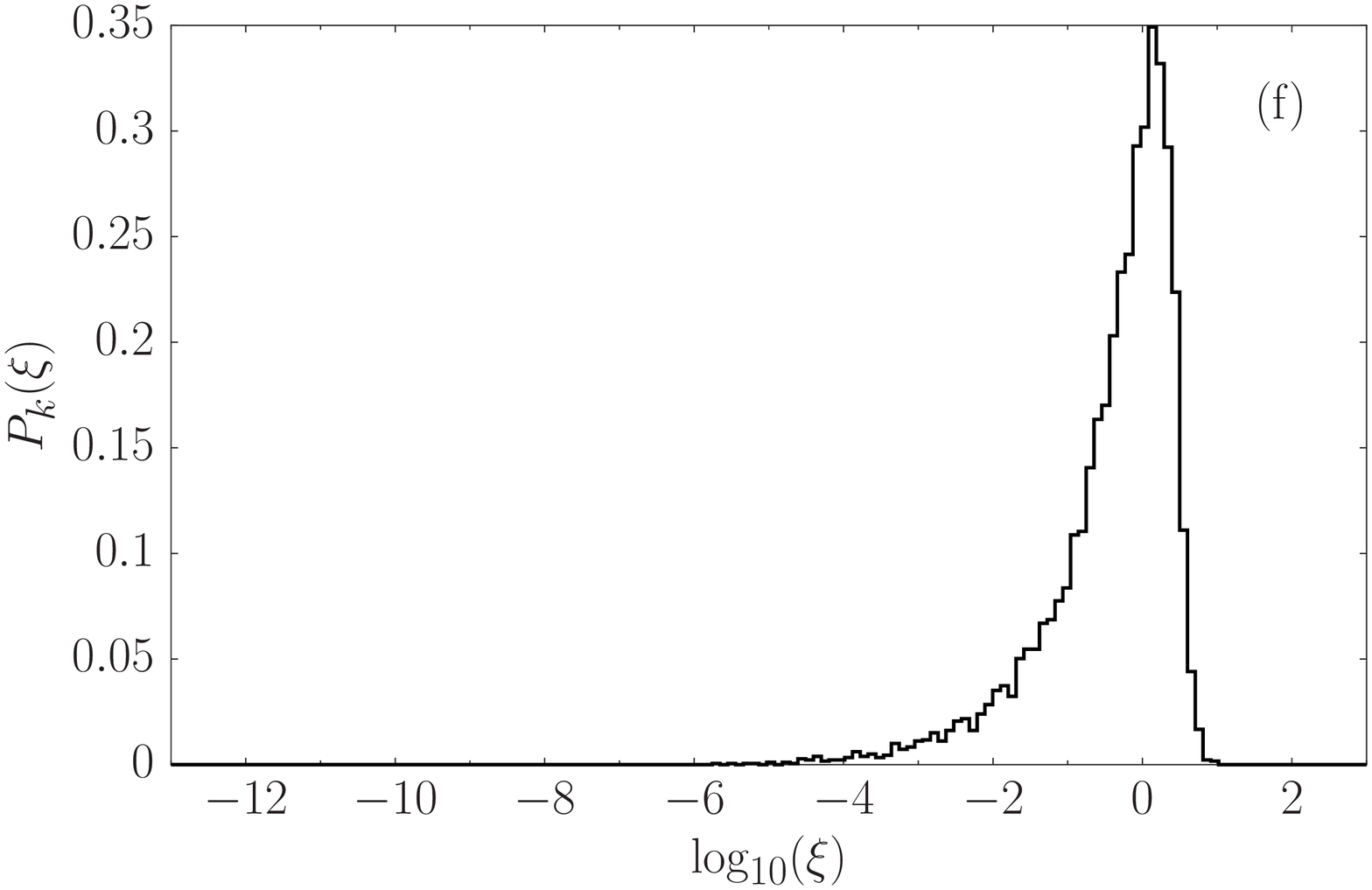}
  \caption{\label{fig7}%
    Distribution $P_k(\xi)$ of the normalized spacings of the
    Shnirelman doublets for (a)~$k=2$, (b)~$k=3$, (c)~$k=250$,
    (d)~$k=750$, (e)~$k=1200$, (f)~$k=1800$.  }%
\end{figure*}

We are interested in the spectral statistics of the spacings of the
energies $E_j$ and $E_{j'}$ of Shnirelman doublets, i.e., the
Shnirelman splittings, each one defines a characteristic tunneling
time given by $T \approx \hbar/\Delta E_j$. Let $E_j$ and $E_{j'}$ be
the energies of a Shnirelman doublet of a specific realization
obtained by diagonalization. We define
\begin{eqnarray}
  \label{xiprime}
  \xi_j' &=& 2\frac{|E_j-E_{j'}|}{|E_j+E_{j'}|},\\
  \label{xi}
  \xi_j &=& \frac{\xi_j'}{\bar{\xi '}},
\end{eqnarray}
where $\bar{\xi '}$ is the mean value of $\xi_j'$ taken over the
corresponding realization. Therefore, the quantity $\xi$ is a measure
of the splitting $E_j$ and $E_{j'}$ in units of the energy of the
doublet, averaged over the splittings of the corresponding realization
of the ensemble. The average is performed in order to compare the
splittings of different realizations of the ensemble; hence,
$\bar{\xi}=1$. Then, $\xi$ is a kind of unfolded spacings of the
Shnirelman doublets.

In ~\Fref{fig7}, we present the distribution of the spacings of
Shnirelman doublets $P_k(\xi)$ for various values of $k$. These
results were obtained for $n=2000$ and $1000$ realizations of the
ensemble.  For $k=2$, this distribution displays a large narrow peak
close to $\xi =1.3\rm{x}10^{-9}$ which is not symmetric. The peak
decays very fast, acquiring a somewhat constant tail toward larger
values of $\xi$; this tail vanishes after $\xi = 100$. That is, while
most doublets have a very small spacing in the normalized units used
here, the doublets close to the edge of the ladder have larger
spacings. For $k=3$, similar results hold; yet, it is worth noticing
that the distribution becomes somewhat wider with respect to the
result for $k=2$. As $k$ increases further, the behavior of $P_k(\xi)$
becomes more complex, with a gradual appearance of a second peak
[close to $\xi=1$ in \Fref{fig7}c]. For a larger value of $k$ around
$k\approx 750$ [see \Fref{fig7}d], both peaks have a similar amplitude;
beyond this value of $k$, the left peak diminishes smoothly, eventually
vanishing, yielding again a single peak distribution, this time for
values centered around $\xi\approx 1$.

The transition in $P_k(\xi)$ described above can be understood as
follows. For small values or moderate values of $k$, the unimodal
distribution reflects the existence of one or more ladders of
Shnirelman doublets. Each ladder has a number of doublets, the spacing
with in each doublet becoming larger (smaller tunneling times) as we
climb up the ladder. Imposing $\bar{\xi}=1$ yields the long tail
observed in the distribution. By increasing the value of $k$, the
distribution $P_k(\xi)$ becomes bimodal, displaying a second peak
centered around $\xi = 1$. That is, the doublets have either a very
small splitting or splittings of order $1$. As mentioned above, larger
splittings are attributed to the last doublets of a ladder around a
stable fixed point. In order to have a significant number of them
without increasing the number of small splittings, we conclude that
their ladder must consist of very few (one or two) Shnirelman
doublets. This idea is consistent with the fact that, for large value
of $k$, only a single doublet is observed around the stable fixed
points where the quantization condition \Eref{eq7} holds. Note that
the latter case implies again a unimodal distribution $P_k(\xi)$, this
time the peak being centered around $1$, as it is observed
numerically.

\section{Summary and conclusions}
\label{Sec:Concl}

In this paper, we have investigated the nearest-neighbor spacing
distribution of the $k$-body embedded ensembles for bosons distributed
in two levels for $\beta=1$ and $\beta=2$. For $\beta=1$, we found a
large peak at $s=0$ in a large interval of $k$ which indicates the
presence of degeneracies. This peak is quite robust in terms of $k$,
disappearing only when $k$ is very close to $n$, the total number of
bosons. For $\beta=2$, the peak is absent, despite of the fact that
there are accidental quasi-degeneracies which yield small spacings;
hence, the large peak is a consequence of the time-reversal invariance
of the ensemble. We showed that this peak is associated with
Shnirelman doublets, which semiclassically correspond to the
quantization of tori that are non-self-retracing under
time-reversal. These results provide further evidence on the
integrability of the ensemble~\cite{BJL2003} based now on the spectral
properties of the ensemble and therefore explain the non-ergodic
properties of the ensemble~\cite{Asaga2001}. The fact that Shnirelman
doublets are not observed for $k$ very close to $n$, where GOE
spectral statistics hold, is due to the fact that the
non-self-retracing tori which would yield such doublets have a
extremely small action, as shown in the phase-space representation of
this case.

We also found for $\beta=1$ that the number of Shnirelman
quasi-degeneracies displays a dependence upon $k$
(cf. \Fref{fig3}). Moreover, the statistics of the normalized
splittings do display also a dependence on $k$; in particular, for
$k=2$ and $k=3$ which are the physically relevant cases, we observe
certain qualitative differences. We believe that these results may be
interesting for understanding and modeling three-body interactions in
Bose-Einstein condensates.

Indeed, the existence of Shnirelman doublets opens the possibility of
producing or observing other type of Josephson-like oscillations in
two-mode Bose-Einstein condensates which may not be centered around
zero population imbalance [see, e.g., Figs.~\ref{fig4}(a) and
\ref{fig4}(b)]. To clarify this, we must emphasize that the degenerate
states present in two-mode Bose-Einstein
condensates~\cite{Salgueiro2007} are not Shnirelman doublets; the
degeneracies are due to the fact that the potential wells are
indistinguishable, i.e., they are associated with the quantization of
two tori related by the symmetry $J \to n+1-J$. Therefore, in order to
produce Shnirelman doublets, we must have the possibility of tuning
{\it all} two-body interaction matrix elements at will, which may
require considering also two species condensates. Once this is done,
the statistical properties of Shnirelman doublets could be used to
characterize the role of interactions beyond $k=2$. This will be the
subject of a future work.

\begin{acknowledgments}
  We have profited from discussions and suggestions with F. Leyvraz,
  C. Jung and T.H. Seligman. We acknowledge financial support from the
  projects IN-107308 (DGAPA-UNAM) and 57334-F (CONACyT). S.H.Q. was
  supported by CONACyT.
\end{acknowledgments}

\end{document}